\documentclass{tran-l}

\usepackage{graphicx}
\usepackage{graphics}
\usepackage{subfigure}

\theoremstyle{definition}

\theoremstyle{remark}

\numberwithin{equation}{subsection}

\begin{document}

\title[Is monogamy of entanglement geometrical?]
 {Is monogamy of entanglement geometrical?}

\author{Xiao Dong, Hanwu Chen, Ling Zhou}

\address{Faculty of Computer Science and Engineering, Southeast University, Nanjing, China}

\email{Xiao.Dong@seu.edu.cn}

%
%
%
%
%
%


\begin{abstract}
  This work aims to understand the monogamy of quantum entanglement from a geometrical point of view. By regarding
    quantum entanglement as a geometrical structure on the state space of quantum systems and attributing all entanglement related properties as emergent from this geometry of entanglement, we assume there exists a genuine general monogamous relation of quantum entanglement w.r.t. a correspondent genuine entanglement measure $Q^*$ which possesses an underlying geometrical origin. We speculate that the monogamous relations w.r.t. an entanglement measure $Q$ can be understood by comparing the different dimension dependencies of the measure $Q$ and $Q^{*}$. We gave evidences of our conjecture by readdressing two observed properties of the monogamy  relations from this geometrical standpoint. Besides the phenomenal explanation of the monogamy of entanglement, we also discussed a fibre bundle structure based candidate solution for the geometry of entanglement and explained how this idea is related to the ER=EPR conjecture and other interesting quantum information processing problems including monogamy of entanglement, entanglement distillation, bound entanglement and activation, and entanglement catalyst.
\end{abstract}

\maketitle

\section{Introduction}
Monogamy of nonclassical correlations such as entanglement and discord constrain their shareability among subsystems of composite systems\cite{Koashi_monogamy_general}. Particularly, among different kinds of nonclassical correlations, the monogamy of entanglement not only plays important roles in quantum information processing tasks such as quantum key distribution, but also is the crucial component of the on-going debate of the AMPS paradox on black hole information\cite{Maldacena_ER_EPR}.  Unfortunately the monogamy of entanglement is still not fully understood.

There are two main obstacles for our understanding of the monogamy of entanglement. The first one is due to our ignorance of the physical nature of entanglement. As a property of entangled systems, monogamy of entanglement should originate from the physical nature of entanglement. Without a complete understanding of the entanglement, it will be difficult to discuss monogamy of entanglement. The second problem is that monogamy relation is defined with respect to a quantum measure. The puzzling situation we face now is that we have to play with a set of inequivalent entanglement measures to have a complete picture of the monogamy of entanglement\cite{Modi_correlation}. We think the current work to check the monogamy relation using different measures is on the wrong track since this will lead to multiple or even an infinite number of versions of monogamy relation since potentially we may define an infinite number of quantum measures.

The guideline of this work is to regard monogamy as an intrinsic signature of entanglement, which originates from the physical nature of entanglement and therefore is valid for all entangled systems. There exists at least one unique entanglement measure which completely reveals the relation between monogamy and the physical nature of entanglement so that the monogamy relation holds for all systems w.r.t. this \emph{genuine monogamy} measure. If this is the case, then we do not need to check the monogamy relation on all the other measures since they are not the right measure for monogamy relation of entanglement. On the other side, all entanglement measures represent certain aspects of the nature of entanglement and therefore they are all related. We can then examine the monogamy properties of different quantum measures by checking their relations with the genuine monogamy measure.

The motivation of our work lies on three papers trying to answer the following questions:
\begin{enumerate}
  \item Are general quantum correlations monogamous?\cite{Streltsov_monogamy_general}
  \item What does monogamy in higher powers of a correlation measure mean?\cite{Geetha_monogamy_higher_power}
  \item Do large number of Parties enforce monogamy in all quantum correlations?\cite{Kumar_monogamy_qubit}
\end{enumerate}

Their conclusions are that:
\begin{itemize}
  \item Any measure of correlations that is monogamous for all states and satisfies reasonable basic properties must vanish for all separable states. It seems that monogamy is a unique characteristic of entanglement and any measure of correlations which is nonzero on some separable states will violate monogamy and therefore excludes the monogamy property for other correlations other than entanglement\cite{Streltsov_monogamy_general}.

  \item All multiparty quantum states can be made monogamous by using positive integral powers of any quantum correlation measure. But such monogamy inequalities are not useful either to quantify the correlations or to show the restricted shareability of correlations in multiparty states\cite{Geetha_monogamy_higher_power}.

  \item There are numerical evidence that almost all pure quantum states of systems consisting of a large number of subsystems are monogamous with respect to all quantum correlation measures of both the entanglement-separability and the information theoretic paradigms, indicating that the volume of the monogamous pure quantum states increases with an increasing number of parties\cite{Kumar_monogamy_qubit}.
\end{itemize}

These observations evoke our curiosity to ask the following questions:
\begin{itemize}
  \item What's the intrinsic nature of entanglement that distinguishes it from other correlations such as the classical correlation or the discord, so that only entanglement can show a general monogamy property? Do the quantum entanglement and the monogamy of entanglement stem from the same root?

  \item Obviously not every quantum measure is compatible with the monogamy relationship in all dimensions. Most efforts have been devoted to examine if the monogamy relationship holds with a given entanglement measure for a certain system. For us we think this is the wrong question since the monogamy should not dependent on the entanglement measure and right ones should be:
      what's the genuine entanglement measure that can fully compatible with the monogamy relationship? Why this genuine measure can hold but other quantum entanglement measures are not monogamous on certain systems?
\end{itemize}

We try to answer these questions by proposing the following hypotheses:
\begin{enumerate}
  \item Entanglement is geometrical.
  \item Entanglement measure is geometrical.
  \item Monogamy of entanglement is geometrical and valid for all systems.
\end{enumerate}

In the left part of this paper we will first explain these hypotheses in details. Then based on our hypotheses we will give our qualitative explanation to the following facts:
\begin{enumerate}
  \item Most entanglement measures do not satisfy the monogamy relationship for general systems but the higher power versions of them can.
  \item The volume of the monogamous pure quantum states increases with an increasing number of parties. In fact according to our hypothesis this is the wrong description since we assume the monogamous property is valid for all pure quantum states. The right description should be:
      For any quantum measure and pure quantum states, with the increase of the number of subsystems, the monogamy relationship with the given measure holds for a larger percentage of all states.
\end{enumerate}
We will also give a general discussion about the geometry of entanglement and the monogamy of entanglement in Section 4 before concluding our work.

\section{Hypotheses}

\subsection{Entanglement is geometrical}

It's is well known that classical correlations are infinitely shareable whereas there is a restriction on the shareability of quantum entanglement amongst the several parts of a multipartite quantum state.  Streltsov\cite{Streltsov_monogamy_general} shows that for quantum-correlations measures that satisfy the following criteria: (a) positivity; (b) invariance under local unitary transformations; and (c) nonincreasing when an ancilla is introduced, monogamy relations cannot hold if the measure does not vanish for separable states. As for quantum discord, although monogamy can still be satisfied for some special cases, it should be guaranteed that monogamy of discord does not hold for general systems due to the result of \cite{Streltsov_monogamy_general}.

What makes quantum entanglement distinctive from other classical and quantum correlations? Why is the monogamy relation valid for quantum entanglement? In fact we do have an entanglement measure which satisfies the monogamy relation in all dimensions, the squashed entanglement. This is a strong hint that monogamy is an intrinsic property of quantum entanglement.

Our assumption to understand the physical nature of quantum entanglement is to regard entanglement as a geometrical structure on the quantum system state space, i.e., quantum entanglement is geometrical.

This assumption does not come from nowhere. Besides the well known basic role played by geometry in such areas of modern physics as general relativity, quantum-field theory, and string theory, in fact there are lots of evidences indicating that a geometrical description is a reasonable way to understand quantum entanglement. Examples are:

\begin{description}
  \item[Building emergent spacetime geometry from entanglement] In \cite{Van_Raamsdonk_spacetime_entanglement} it's argued that quantum entanglement appears to be crucial for the emergence of classical spacetime geometry. Classical spacetime can be build up by entangling degrees of freedom and tear them apart by disentangling. Swingle shows that holographic spacetimes can be constructed using entanglement renormalization\cite{Swingle_spacetime_entanglement}. The ER=EPR conjecture\cite{Maldacena_ER_EPR} says that the quantum degrees of freedom corresponding to black holes connected by an Einstein-Rosen (ER) bridge are entangled. The entanglement entropy is related to the cross sectional area of the bridge and conversely all entanglement between causally disconnected degrees of freedom are geometrized by an ER bridge. \cite{Hensen_EPR_wormhole}\cite{Lobo_EPR_wormhole}\cite{Gharibyan_ER_EPR_inequation} show evidences for the ER=EPR conjecture. The consistency between ER=EPR and quantum mechanics was discussed in \cite{Susskind_ER_EPR}. In \cite{Susskind_ER_bridge_nowhere}\cite{Susskind_ER_bridge} Susskind indicates that both the entanglement and the complexity of quantum systems play essential roles in the evolution of black holes and EPR bridges. Due to the fact that entanglement is so closely related with the spacetime geometry, it's natural to assign quantum entanglement with a geometrical nature. But the analytical picture between entanglement patterns and ER bridges is still missing.

  \item[Understanding entanglement with geometric tools] Efforts to understand entanglement from a geometrical point of view also go back to the work of \cite{Mosseri_entanglement_Hopf}\cite{Levay_entanglement_connection}. Among them the most interesting work is to understand 2 qubit pure state system from Hopf fibration, where the entanglement is understood as the twisting of the fibre which is defined by the natural connection on the Hopf fibration. For more details please refer to \cite{Levay_entanglement_connection} and related works\cite{Bernevig_entanglement_connection}\cite{Levay_entanglement_twistor}\cite{Heydari_entanglement_general}. This provides an alternative approach to understand entanglement from a geometrical standpoint.

  \item[Unified view of quantum and classical correlations by geometric measures] In \cite{Modi_unified_correlation}\cite{Bromley_correlation_Bures} a unified approach was addressed to understand quantum and classical correlations, in which different correlations including quantum entanglement, discord, dissonance and total correlation, are quantified in a geometrical manner using either relative entropy or Bures distance. What's more for two qubit systems, the important concept of concurrence in understanding quantum entanglement is closely related with the geometrical concept of Bures distance of the state space as indicated in \cite{Streltsov_entanglement_Bures}. This also indicates the connection between entanglement and geometrical concepts.

   \item [Holographic entanglement entropy (HEE) and area laws in quantum system]  Entropy and mutual information are closely related entanglement in that they help to build important entanglement measures such as the entanglement of formation and squashed entanglement. \cite{Nishiokaa_HEE}\cite{Wolf_area_law} introduced the recent progresses on the holographic understandings of the entanglement entropy in the AdS/CFT correspondence. They also showed that both the entanglement entropy and mutual information posses an important geometrical property, the area law, which means both the entanglement entropy and mutual information are proportional to the area of the minimal surfaces separating the subsystems in the dual space of the system. This means that both entropy and mutual information have their geometrical pictures. These works show the quantitative relation between entanglement and geometry.

\end{description}

From the above observations, we may claim that entanglement is closely connected with geometry, although we are not sure for now if entanglement should be correspondent to a fibre bundle or other geometrical structures.

If the physical nature is essentially a geometrical structure, we may conjecture that all properties of entanglement, including different entanglement measures and the monogamy relation, are emergent from this structure.

A crucial property of this idea is that the geometrical structure is dimension dependent. That's to say, for different systems with different state space dimensions, we may have different geometrical structures for the entanglement. For example we do not have SLOCC inequivalent entangled systems for 2 qubit pure states but there exist SLOCC inequivalent entangled states for more than 3 qubits \cite{Horodecki_entanglement}. If the entanglement does correspond to the twisting of the fibre bundle on the state space, the SLOCC inequivalence can then be regarded as different fibre bundles or inequivalent connections on the state space. We may further extend this idea to attribute other observed properties of entanglement, such as bound entanglement and its activation, entanglement catalyst, to the dimension dependency of the geometrical structure of entanglement as will be mentioned in the discussion session.

Though there are signs that a geometrical picture might be only valid for entanglement but not for other correlations such as discord\cite{Streltsov_monogamy_general}, we cannot completely exclude the possibility that discord also possess a correspondent geometrical structure. The result of \cite{Streltsov_monogamy_general} may only mean that the possible geometrical structure of discord does not imply the existence of a general monogamy relation of discord.

Hypothesis 1: Quantum entanglement originates from a geometrical structure on the state space of entangled systems.

\subsection{Entanglement measure is geometrical}
One main puzzle about entanglement is that there are multiple entanglement measures which are not equivalent to each other \cite{Vedral_entanglement}. If entanglement has a geometrical nature, then different quantum information tasks explore different aspects of this structure and accordingly different quantum measures partially reveal different properties of the structure. In fact, one of the key characteristics of entanglement measure, that the entanglement is invariant under local unitary transformations, already give a strong hint that entanglement measure is geometrical.

Generally there are two main categories of correlation measures in quantum systems: geometric measures and information theoretic measures. Roughly they quantify correlations by certain minimal geometrical distances and entropy based formulae respectively\cite{Vedral_entanglement}\cite{Modi_unified_correlation}.
Both of them have their correspondent geometrical pictures\cite{Modi_correlation}. We also note that although in the holographic entanglement entropy scheme there are area laws for both the entanglement entropy and mutual information\cite{Nishiokaa_HEE}, we do not know if we have the same geometrical property for the entropy and mutual information used in the entanglement measure. But we assume a similar situation may also apply.

From this geometrical point of view, for a bipartite entangled system $\rho_{AB}$, roughly the geometric measures define the entanglement as the \emph{distance} between the state $\rho_{AB}$ and its nearest separable state; while the information theoretic measures define the entanglement as the area of the minimal surface separating the two subsystems A and B in a certain dual space determined by the state space of $\rho_{AB}$.

Till now we do not have a complete understanding of the relationship between these two categories of quantum measures. But we may claim that

\begin{itemize}
  \item These two types of entanglement measures are correlated.
  \item They have different dimension dependencies.
\end{itemize}

A concrete example to show the correlation between them can be carrid out by entanglement measures of 2 qubit systems. According to \cite{Levay_entanglement_connection}, for a 2 qubit pure state with its Schmidt decomposition given by $\psi_{AB}=\sqrt{\lambda}|\psi_{A}>|\psi_{B}>+\sqrt{1-\lambda}|\psi_{A}^{\perp}>|\psi_{B}^{\perp}>, \lambda>\frac{1}{2}$, its distance to its nearest separable state is $D_{FS}=acos(\sqrt{\lambda})$ and the concurrence $\mathcal{C}$ is given by $\mathcal{C}=2\sqrt{\lambda(1-\lambda)}$. For mixed 2 qubit states $\rho_{AB}$, $\mathcal{C}$ is closely related with the Bures distance between $\rho_{AB}$ and its nearest separable state as $D_{B}(\rho)=2-2\sqrt{(1+\sqrt{1-\mathcal{C(\rho)}^{2}})/2}$\cite{Streltsov_entanglement_Bures}. So there is a closed form relation between $\mathcal{C}$ and the geometric measures $D_{FS},D_{B}$. On the other hand it's well known for 2 qubit systems, the entanglement of formation $E_{F}$ is a function of $\mathcal{C}$ given by $E_{F}(\rho)=h((1+\sqrt{1-\mathcal{C}(\rho)^{2}}/2))$ with $h(x)=x\log_{2}(x)-(1-x)\log_{2}(1-x)$. While $D_{FS},D_{B}$ are examples of distance based geometric measures and $E_{F}$ belongs to the information theoretic measures. Obviously these two types of entanglement measures are connected through $\mathcal{C}$.

For the evidence of the different dimension dependencies of these two categories of quantum measures, a direct observation is that they have different geometrical pictures, i.e., they define the entanglement by 1 dimensional distance and higher dimensional minimal surface respectively (if we assume that some kind of area law is also valid for information theoretic measures). More generally, since the geometrical structure of entanglement is dimension dependent, quantum measures that originate from this geometrical structure should also be dimension dependent. In \cite{Stanford_complexity} is was suggested that the relation between the state complexity, which should be dimension independent from the quantum circuit point of view, and the ER bridge volume, which is related to the entanglement, is dimension dependent. This also provides some hints that our statement on the dimension dependency of entanglement measure is reasonable.

\begin{figure}
  \centering
  \includegraphics[width=10cm]{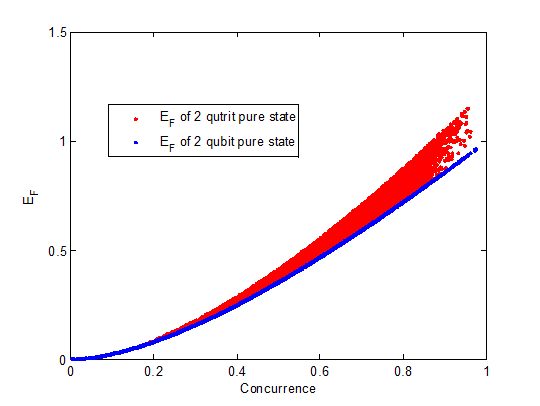}
  \caption{Relation between concurrence $\mathcal{C}$ and entanglement of formation $E_{F}$ for 2 qubit and 2 qutrit systems by numerical simulation}\label{fig_c_eof}
\end{figure}

An intuitive example is to check the most commonly used measures $\mathcal{C}$ and $E_{F}$ for different systems. For 2 qubit pure states $\psi_{AB}$, we know that both $\mathcal{C}$ and $E_{F}$ are completely determined by reduced density matrix $\rho_{A}$ or $\rho_{B}$, and $\rho_{A}$ or $\rho_{B}$ is a complete representative of the entanglement\cite{Fortin_partial_trace}. For 2 qubit mixed states $\rho_{AB}$, $\mathcal{C}$ and $E_{F}$ can only depend on $\rho_{AB}$ but not on $\rho_{A}$ or $\rho_{B}$. In both cases, there is a fixed relation between $\mathcal{C}$ and $E_{F}$. As we go to 2 qutrit pure states, the fixed relation between $\mathcal{C}$ and $E_{F}$ is then lost as shown in Fig. \ref{fig_c_eof}. So both $\mathcal{C}$ and $E_{F}$ can only represent partial property of the geometric structure of entangled states and they are dimension dependent. More detailed discussion on the dimension dependency of quantum measures will be given later.

%

Hypothesis 2: Different quantum entangle measures reveal different aspects of the geometrical structure of the quantum entanglement and generally different quantum entanglement measures have different dimension dependencies.

\subsection{Monogamy is geometrical}
Generally for a composite system the monogamy relation is given by
\begin{equation}\label{eq_monogamy}
  Q(1:2...n)\geq \sum_{i=2}^{n} Q(1:i)
\end{equation}
 where $Q$ is an correlation measure \cite{Modi_correlation}.

As pointed out by \cite{Streltsov_monogamy_general}, the monogamy relation can only hold in general for correlation measures that vanish on separable states.

It's natural to conjecture that general monogamy relation is only possible for entanglement and therefore monogamy is intrinsically encoded in the geometry of entanglement. Geometrically the monogamy inequation represents the relation of certain geometrical parameters, i.e. distances and areas of minimal surfaces for geometric and information theoretic measures respectively. An intuitive geometric picture of the monogamy relation can be that the area of the minimal surface between subsystem 1 and the complementary subsystem including subsystems 2 to n is bigger than the sum of the areas of the minimal surfaces between subsystem 1 and each individual subsystem 2 to n for a information theoretic measure as the squashed entanglement.

We know that for all entanglement measures except for the squashed entanglement, the monogamy relation is only valid for certain special cases, for example for qubit systems monogamy relation holds for $C^{2}$ and square of negativity but not for $E_{F}$\cite{Streltsov_monogamy_general}. So we conjecture that there is at least one intrinsic monogamy of entanglement which is valid for all entangled states, the correspondent genuine entanglement measure is $Q^{*}$ and the squashed entanglement is a candidate for $Q^{*}$.

The importance of proposing the intrinsic monogamy of entanglement is that since it's supposed to hold for all entangled states, it sets general constraints on the valid configurations of the system. Since both $Q^{*}$ and a general quantum measure $Q$ are determined by the same geometrical structure, they may be related by a function $Q=f(\rho,Q^{*})$ for a certain system $\rho$. Then the general monogamy relation w.r.t. $Q^{*}$ can be used to investigate the monogamy relation w.r.t. $Q$.

Unfortunately the only candidate for the genuine measure $Q^{*}$, squashed entanglement, is very difficult to compute\cite{Koashi_monogamy_general}. So the above mentioned idea has little practical quantitative usage. But qualitatively we can explore this idea to understand the monogamy relation for general quantum measure $Q$.

The key ingredient of our hypothesis here is that the genuine entanglement measure $Q^{*}$ should be dimension dependent on the state space of the system, since the geometrical nature of the monogamy relationship asks for it so that it should be valid for all systems with different dimensions . And all the other entanglement measures that do not show the same dimension dependent characteristic, such as the concurrence and entanglement of formation, can not be fully compatible with the monogamous relationship in all dimensions so that the monogamous relationship only holds for a limited number of situations using these measures \cite{Kumar_monogamy_qubit}\cite{Kim_monogamy_general}.

The reason for our assumption of the dimension dependent genuine measure for monogamy of entanglement lies in the following considerations:
\begin{itemize}
  \item If the squashed entanglement defined using entropy and mutual information is the genuine measure for monogamy relation, then it might be dimension dependent just as in the holographic entanglement entropy case.
  \item If the monogamy relation originates also from the geometrical structure of the entangled state space, then it's highly possible the monogamy relation is dimension dependent. For example if we think the genuine measure here is really correspondent to the minimal surface area between subsystems, then the monogamy relation can be explained geometrically as: the area of the minimal surface between subsystem 1 and 2:n should be larger than the sum of the areas of the minimal surfaces between subsystem 1 and each individual subsystems 2 to n. Obviously since the left/right side of the monogamy relation involve systems with different dimensions, then the genuine measure should be dimension dependent.
   \item The existing observations on the monogamy of entanglement \cite{Geetha_monogamy_higher_power}\cite{Kumar_monogamy_qubit} show a highly dimension sensitive characteristic.
\end{itemize}

Hypothesis 3: The monogamy relation of entanglement also has a geometrical origin and there exist at least one intrinsic monogamy relation valid for all entangled systems. The genuine entanglement measure for this property is dimension dependent.

\section{Understanding monogamy with geometry}
In this part we will use the above given hypotheses to re-address two of the three questions highlighted at the beginning of the paper.
\begin{itemize}
  \item Q1: What does monogamy in higher powers of a correlation measure mean?
  \item Q2: Do large number of parties enforce monogamy in all quantum correlations?
\end{itemize}

From now on our discussion will focus on the monogamy of the entanglement of multiple qubit pure states.

\subsection{Dimension dependency of entanglement measures}
By our assumption, the monogamy of entanglement $Q^{*}(1:2...n)\geq \sum_{i=2}^{n} Q^{*}(1:i)$ holds for all entangled systems and $Q^{*}$ is the dimension dependent genuine measure for monogamy, which represents the geometrical nature of the monogamy of entanglement.

By the concept of dimension dependency we mean that the quantum measure $Q$ is sensitive to the dimension of the system concerned.
In order to quantify the dimension dependency, we need to introduce the concept of a characteristic \emph{scale} $\varepsilon \in [0,1]$ ,$\varepsilon=0$ and $\varepsilon=1$ for separable and maximal entanglement respectively, which is can be regarded as a dimension independent quantum measure, as a common reference. For example $\varepsilon$ can be taken as one of the geometric quantum measures which is essentially a one dimensional \emph{distance} such as the normalized Fubini-Study distance between an entangled 2 qubit pure state and its nearest separable state. Then for a n-qubit system,each quantum measure $Q$ between the first qubit and the rest qubits may be represented as a function as $Q(1:2...n)=q(n,\varepsilon)$ if we assume all different quantum measures are all encoded in the same geometrical structure.

Quantitatively we can introduce a highly simplified version of the dimension dependency as $Q(1:2...n)= \varepsilon^{s_{Q}(n)}$ and \ $s_{Q}(n)$ is monotonically non-decreasing with $n$. The non-decreasing assumption is based on the fact that with a larger $n$, we are dealing with systems with higher dimension and complexity. So any reasonable entanglement measure $Q$ should be more complex. Of course this does not work for practical situation and also it's not possible to get an analytical expression of $s_{Q}(n)$. But we will see such a simplified model can simplify our discussion and help us to grasp the geometrical picture of the monogamy relation.

\begin{figure}
  \centering
  \includegraphics[width=12cm]{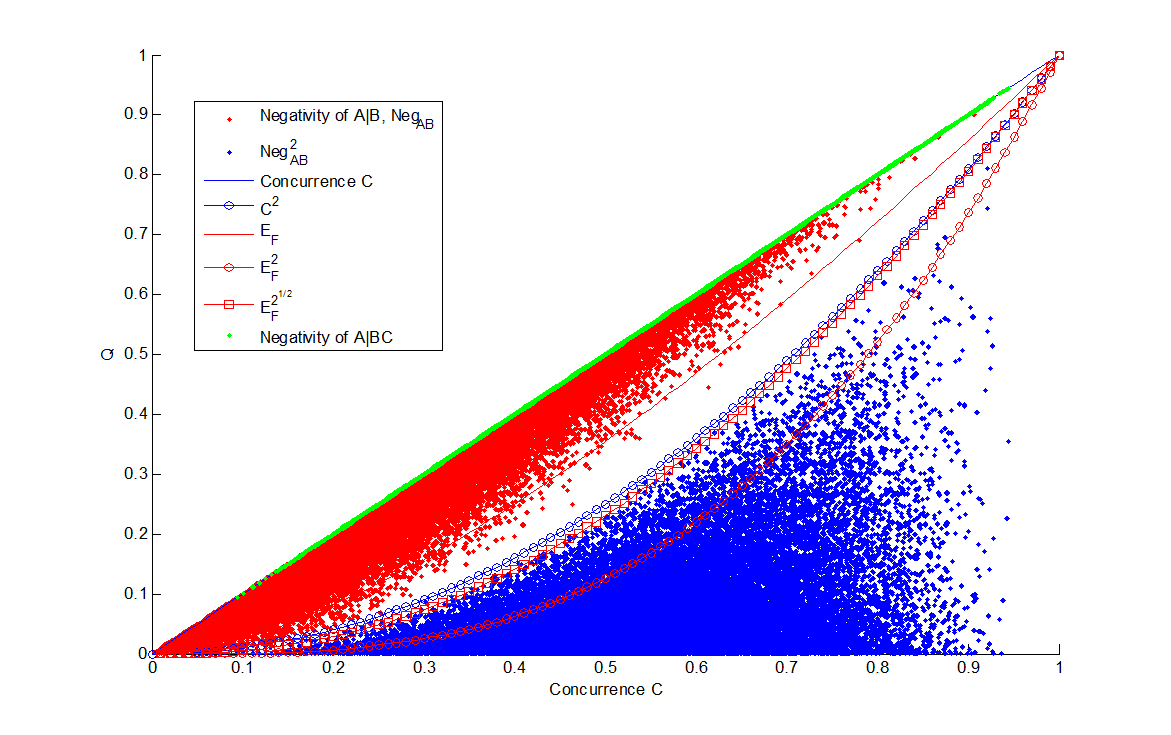}
  \caption{Dimension dependencies of different quantum measures in a 3 qubit pure state system taking $\mathcal{C}$ as the common reference. $s_{\mathcal{C}}(2)=1$, $s_{\tau}(2)=2$, $1<s_{E_{F}}(2)<2$, $s_{E_{F}}(2)>2$, $s_{Neg}(3)=1$ and $1<s_{Neg}(2)<2$.}\label{fig_dependency}
\end{figure}

As we mentioned before, concurrence $\mathcal{C}$ seems to have a dimension of one (distance) for our 2 qubit systems. and we can take concurrence to show the meaning of the dimension dependency. For a general entanglement measure $Q$, the dimension dependency is then defined as $Q(1:2...n)=q(n,\mathcal{C})$ for an entanglement measure $Q$. The dimension dependency of different quantum measures on a 3 qubit pure state system is shown in Fig. \ref{fig_dependency}. Obviously we can see that concurrence $\mathcal{C}$ and the tangle $\tau=\mathcal{C}^{2}$ are dimension independent entanglement measure with respect to $\mathcal{C}$, while $E_{F}$ and negativity are generally dimension dependent. We also see that for negativity there is no fixed value for $s_{Neg}(2)$, but still we can set a range of $s_{Neg}(2)$ so that our simplified model can still work by checking its properties with its upper/lower limits. Also we can observe that there do exist different dimension dependencies for different measures.
Of course we need to point out that in fact $\mathcal{C}$ should not be a good candidate for the reference since $\mathcal{C}$ itself is dimension dependent. We can see this by the fact that for pure qubit systems, there is a fixed relation between $\mathcal{C}$ and $E_{F}$, and we are pretty sure that $E_{F}$ is dimension dependent. Here we use $\mathcal{C}$ as the reference only to show the existence of dimension dependency and our simplified model is reasonable.

Now if we go back to assume that we do find a proper $\varepsilon$, we have that for a given quantum measure $Q$, the monogamy relation of a qubit system is given by
\begin{equation}\label{eq_monogamy_simplified}
  \varepsilon(1:2...n)^{s_{Q}(n)}\geq \sum_{i=2}^{n} \varepsilon(1:i)^{s_{Q}(2)}
\end{equation}

Fig. \ref{fig_all} shows the boundaries between monogamous and non-monogamous configurations for a 3 qubit pure state system using different measures also using $\mathcal{C}$ as the common reference.

Now we will use our simplified model of dimension dependency to re-state the observations about monogamy of entanglement in [][].

\subsection{Q1: What does monogamy in higher powers of a correlation measure mean?}

It's well known that tangle $\tau$ and $E_{F}^{2}$ satisfy the monogamy relation for pure 3 qubit system but concurrence $\mathcal{C}$ and $E_{F}$ do not\cite{Kumar_monogamy_qubit}. The problem that all multiparty states can be made monogamous by considering higher integral powers
of a non-monogamous quantum correlation measure was discussed in \cite{Geetha_monogamy_higher_power}\cite{Kumar_monogamy_qubit} with both analytical and numerical verifications.

We will interpret this observation from our geometrical understanding of quantum entanglement. We show that this is exactly a sign of different dimension dependencies of different quantum measures.

In order to analysis the dimension dependencies of quantum measures, we choose to use concurrence $\mathcal{C}$ as the reference. As we already mentioned, $\mathcal{C}$ itself is not dimension independent, i.e., $\mathcal{C}(1:2...n)=\varepsilon(1:2...n)^{s_{\mathcal{C}}(n)}$ for a n-qubit system. It seems that $\mathcal{C}$ is not a proper reference. But since we are now deal with a certain fixed system, for example a n-qubit pure state, the monogamy relation of a measure $Q$ is given by \ref{eq_monogamy_simplified}. So even $\mathcal{C}$ is dimension dependent, putting $\mathcal{C}(1:2...n)=\varepsilon(1:2...n)^{s_{\mathcal{C}}(n)}$ in \ref{eq_monogamy_simplified} we have the monogamy relation given by
$\mathcal{C}(1:2...n)^{\frac{s_{Q}(n)}{s_{\mathcal{C}}(n)}}\geq \sum_{i=2}^{n} \mathcal{C}(1:i)^{\frac{s_{Q}(2)}{s_{\mathcal{C}}(2)}}$, where both $s_{\mathcal{C}}(n)$ and $s_{\mathcal{C}}(2)$ are constants and do not essentially influence our analysis.

\begin{figure}[tbp]
  \centering
  \mbox{
  \subfigure[Concurrence $\mathcal{C}$,$s_{Q}(2)=1, s_{Q}(3)=1$]
  {\includegraphics[width=4cm]{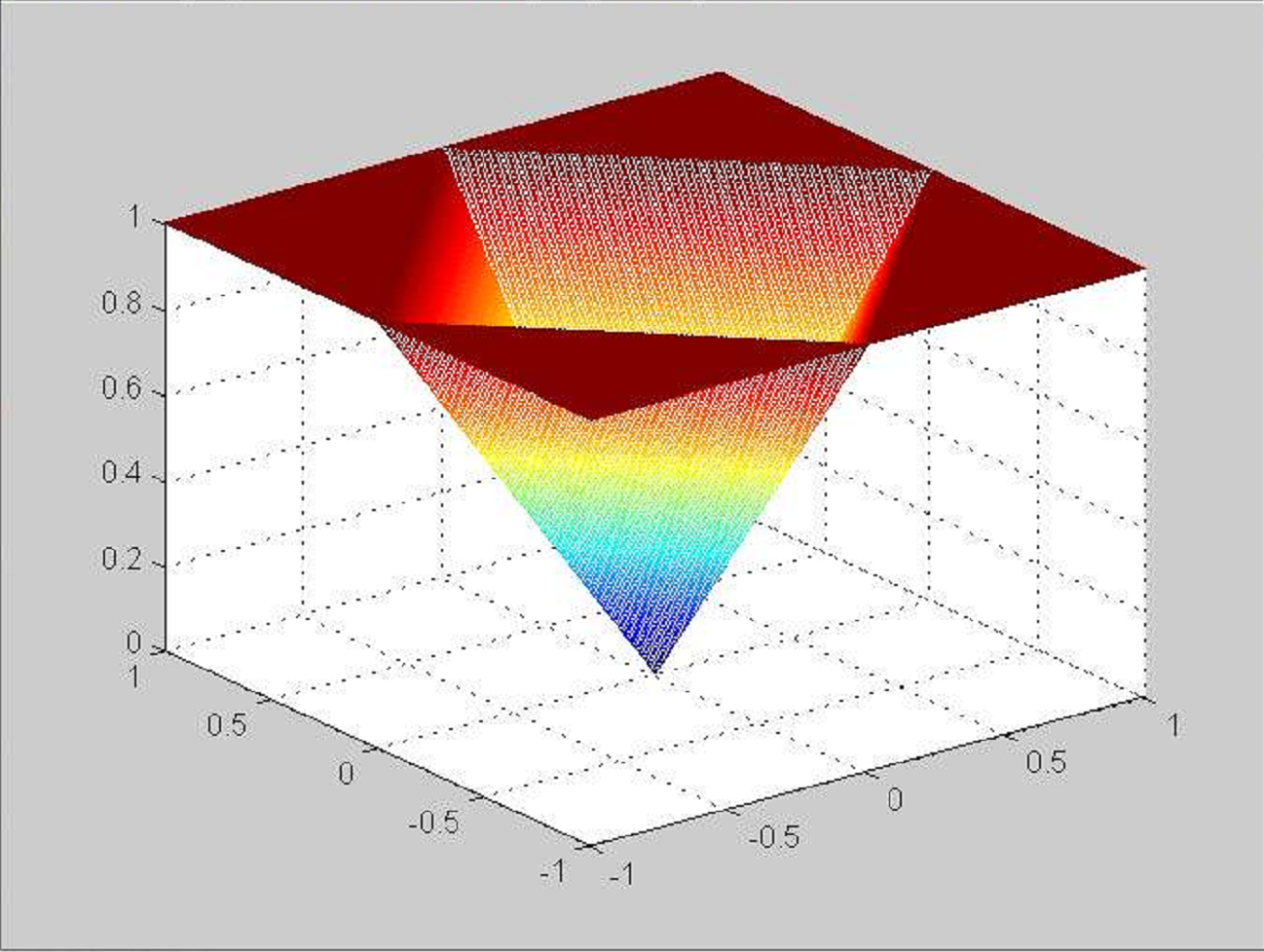}
  \label{fig_111}}

  \subfigure[Tangle $\tau$, $s_{Q}(2)=2, s_{Q}(3)=2$]{
    \includegraphics[width=4cm]{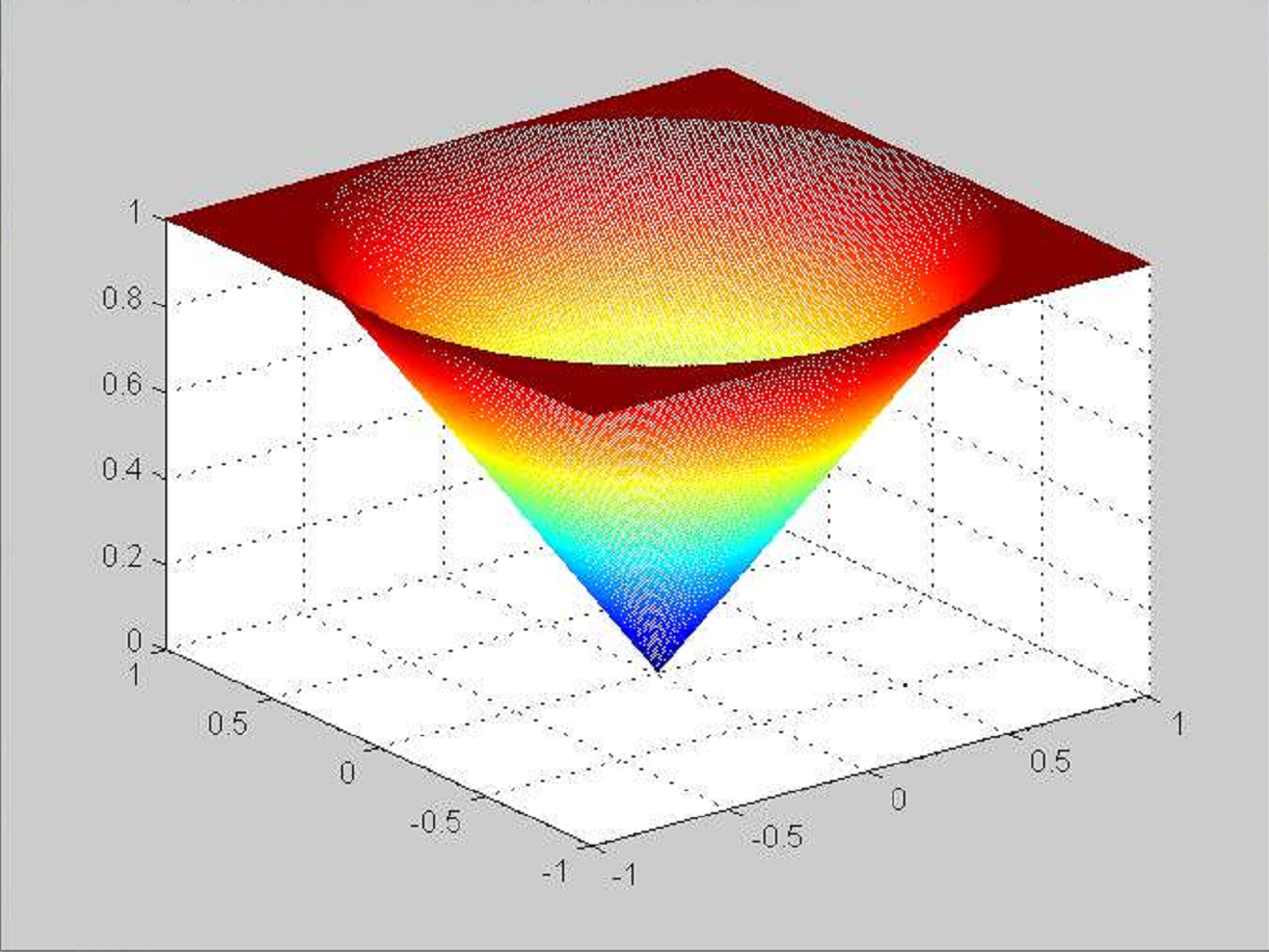}
  \label{fig_222}}

  \subfigure[$s_{Q}(2)=2, s_{Q}(3)=4$]{
    \includegraphics[width=4cm]{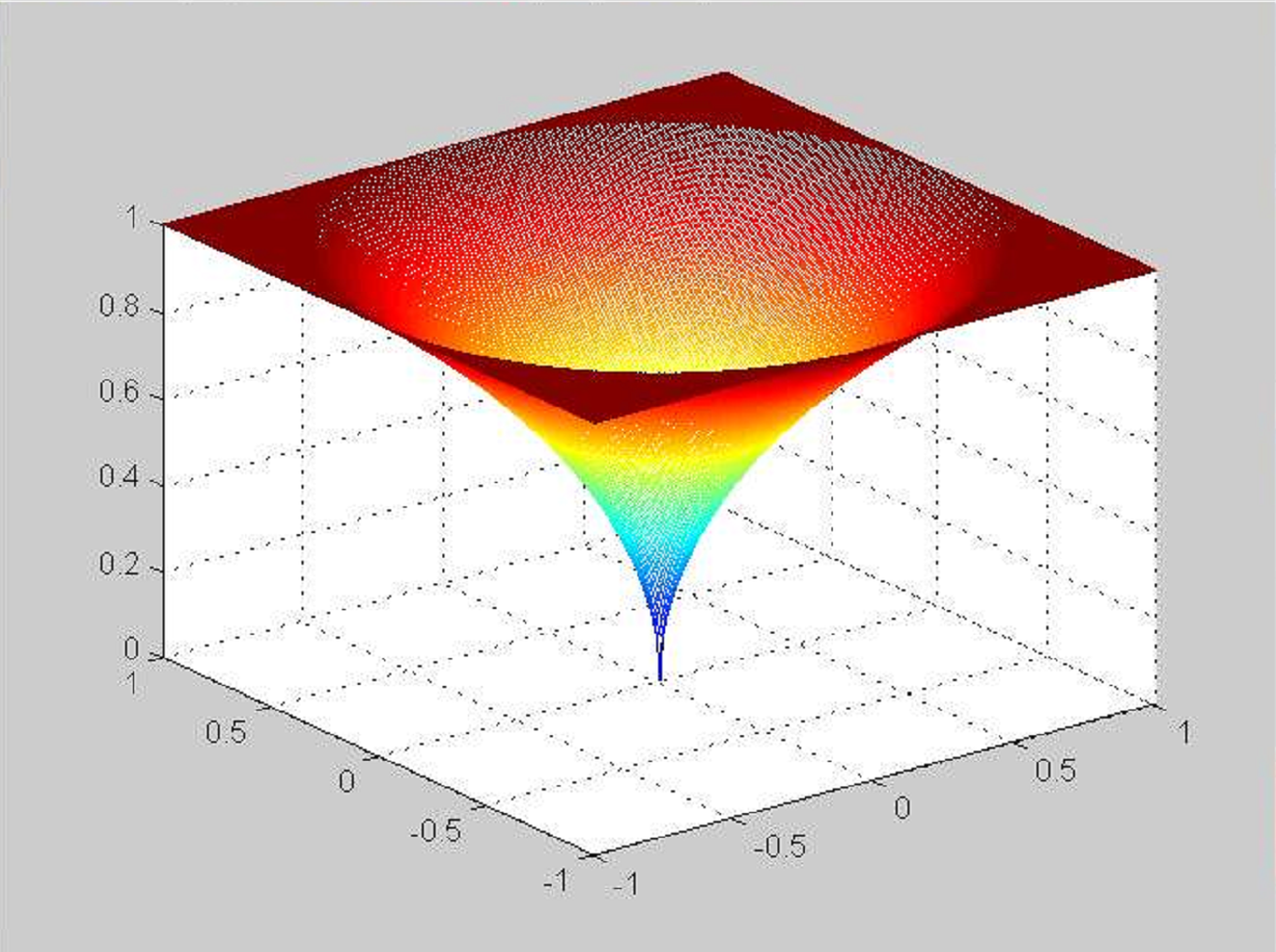}
  \label{fig_422}}}
  \vspace{0.2cm}

  \mbox{
  \subfigure[$s_{Q}(2)=2, s_{Q}(3)=8$]
  {\includegraphics[width=4cm]{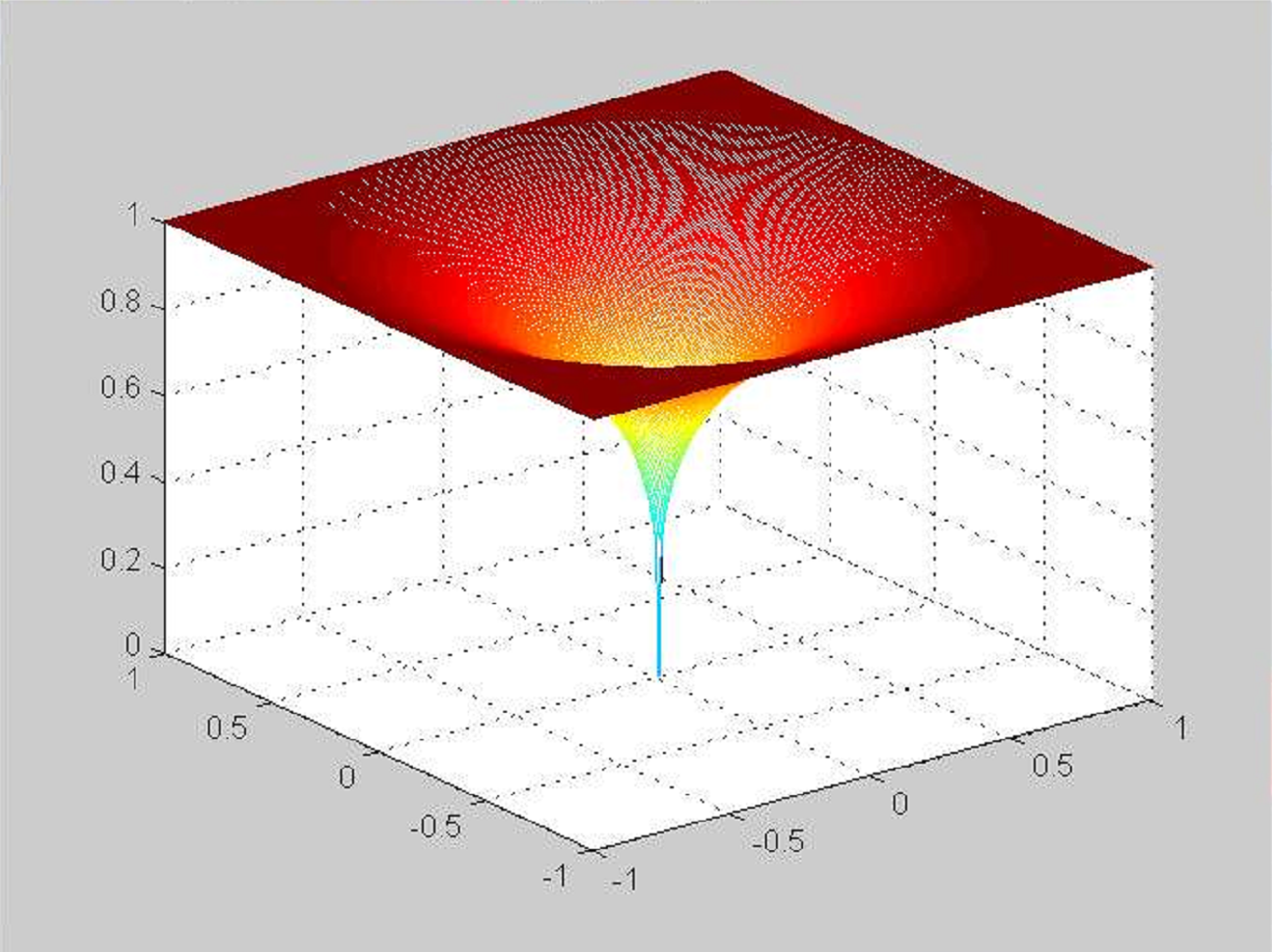}
  \label{fig_111}}

  \subfigure[$s_{Q}(2)=4, s_{Q}(3)=4$]{
    \includegraphics[width=4cm]{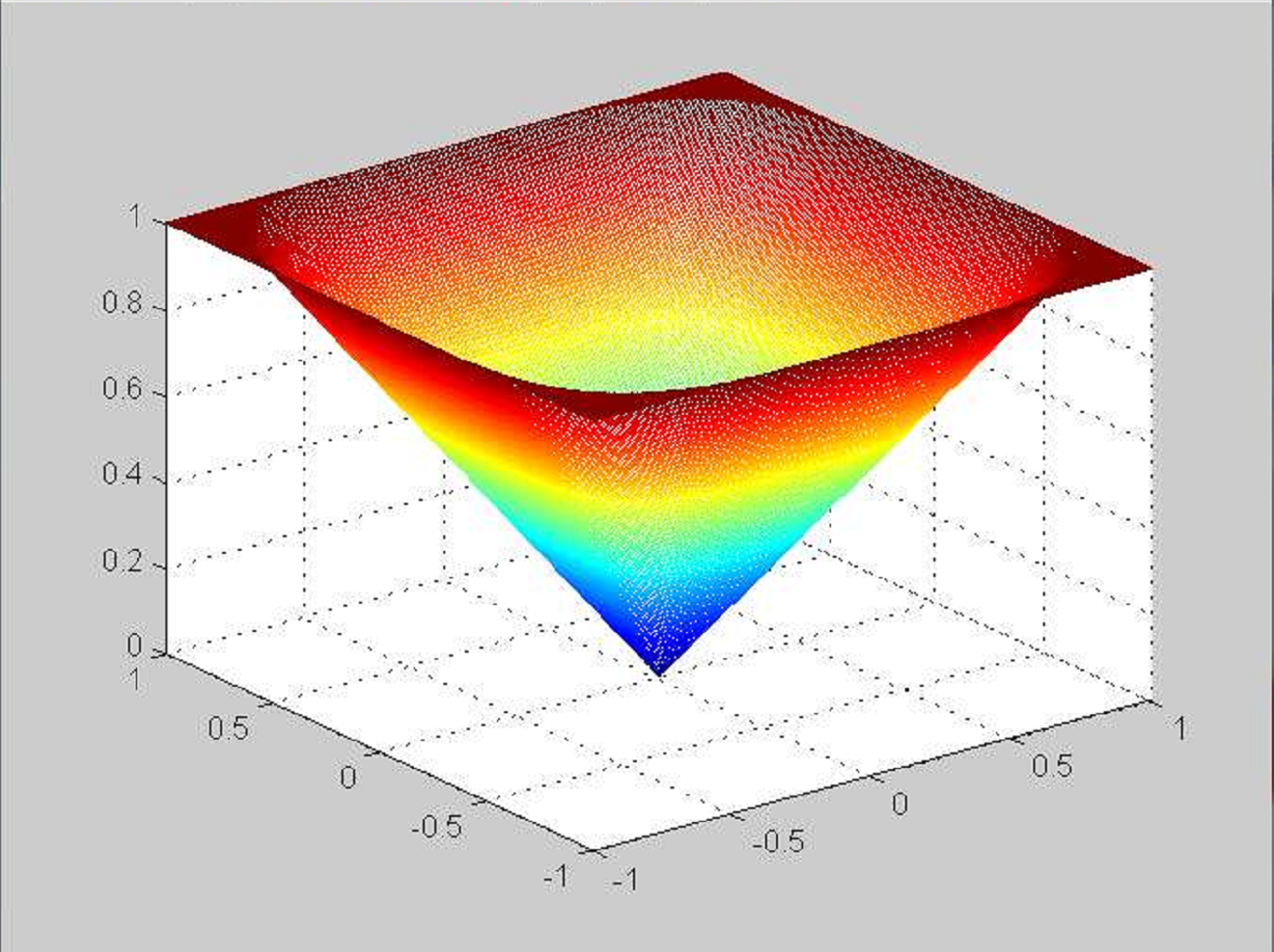}
  \label{fig_222}}

  \subfigure[$s_{Q}(2)=4, s_{Q}(3)=8$]{
    \includegraphics[width=4cm]{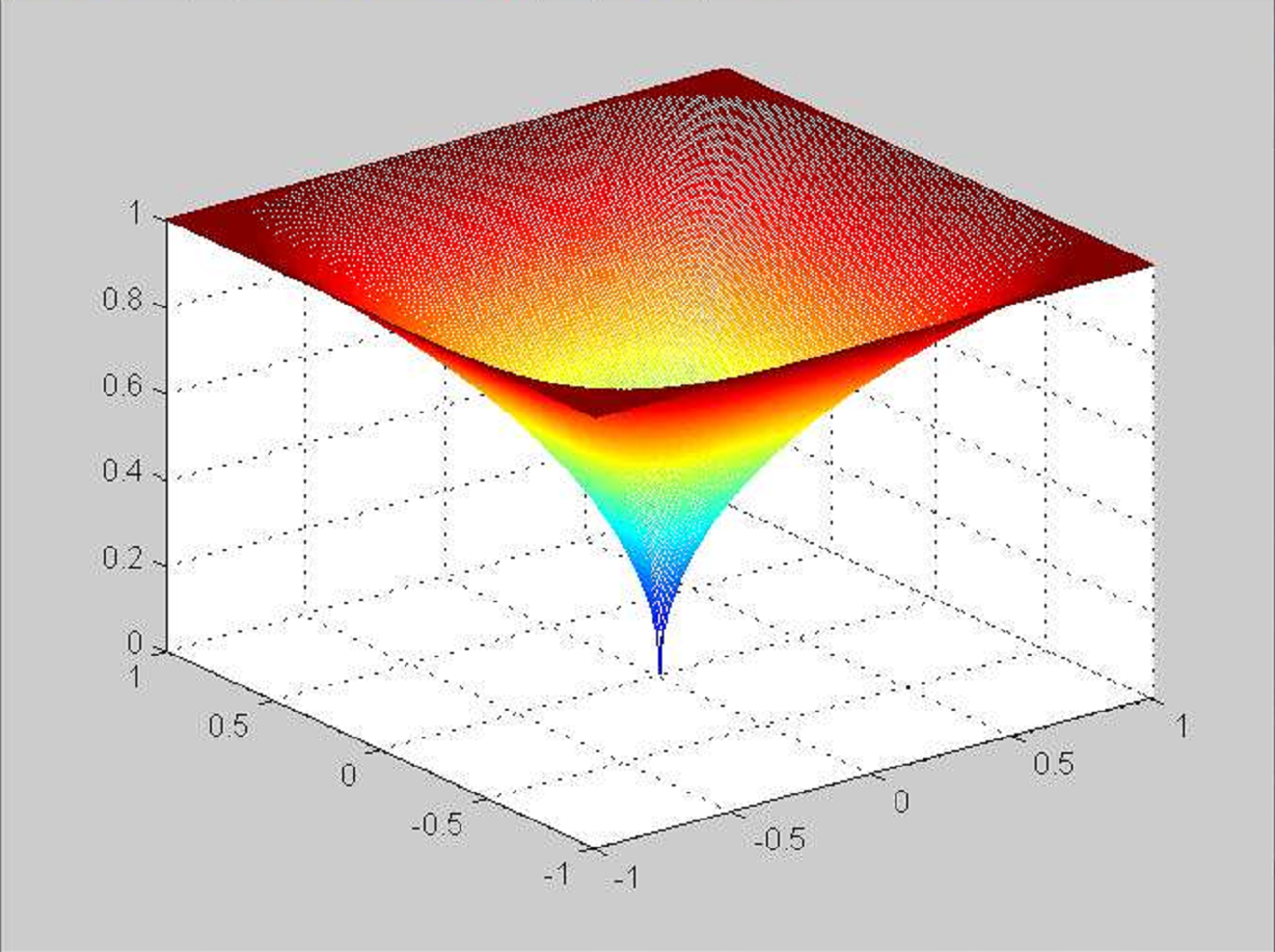}
  \label{fig_422}}}
  \vspace{0.2cm}

  \mbox{
  \subfigure[$s_{Q}(2)=8, s_{Q}(3)=8$]
  {\includegraphics[width=4cm]{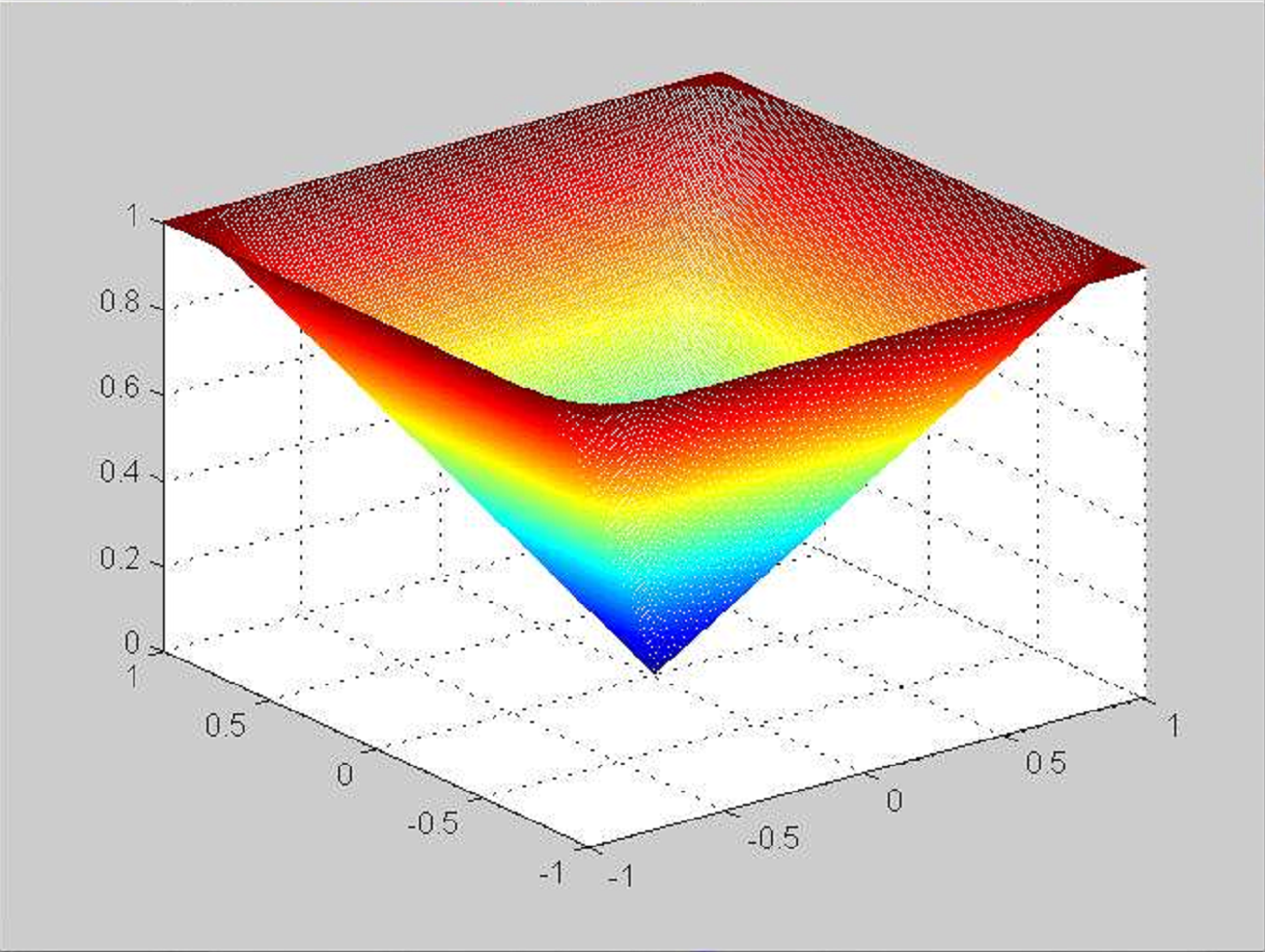}
  \label{fig_111}}

  \subfigure[$E_{F}$]{
    \includegraphics[width=4cm]{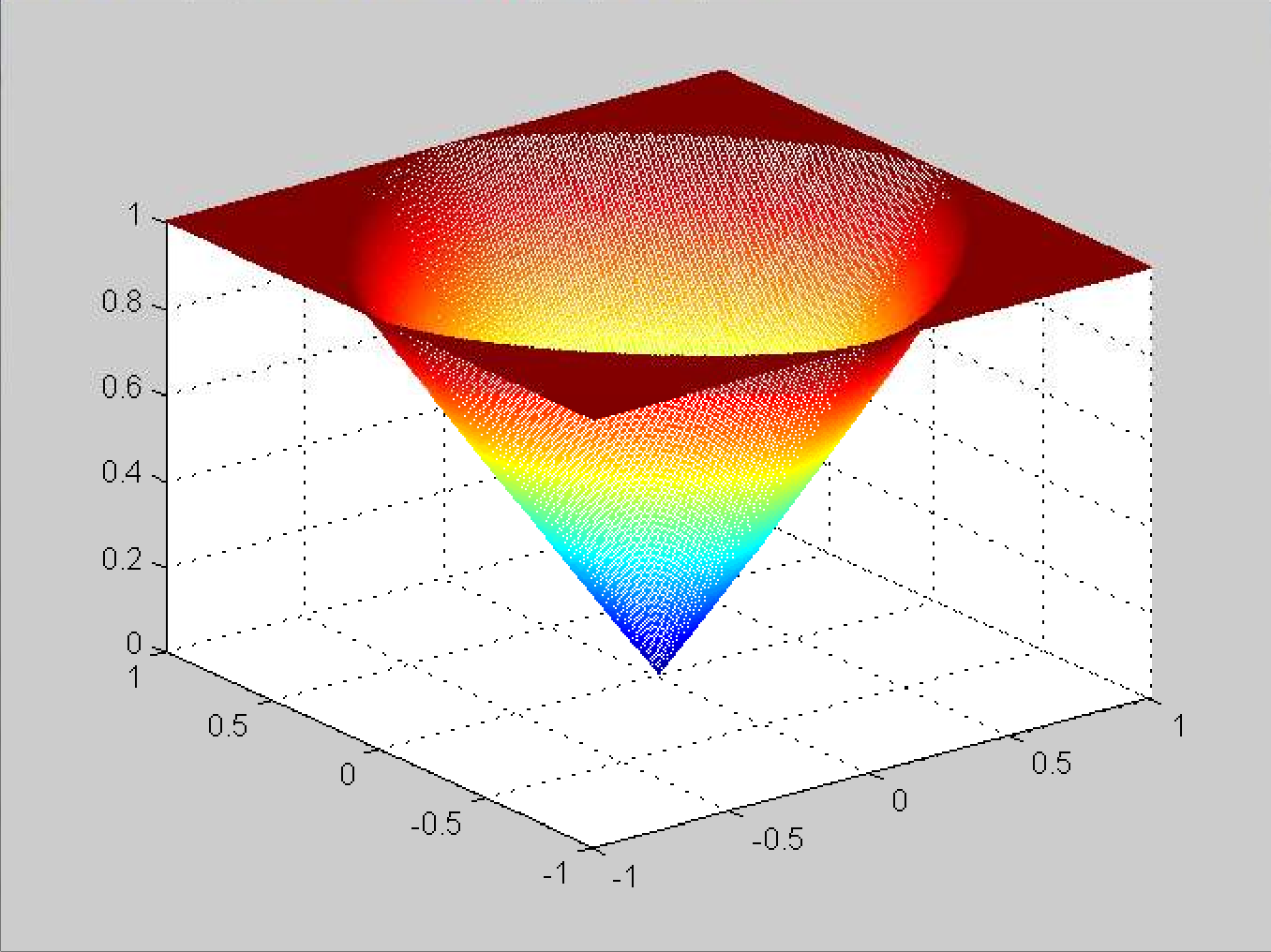}
  \label{fig_222}}

  \subfigure[$E_{F}^{2}$]{
    \includegraphics[width=4cm]{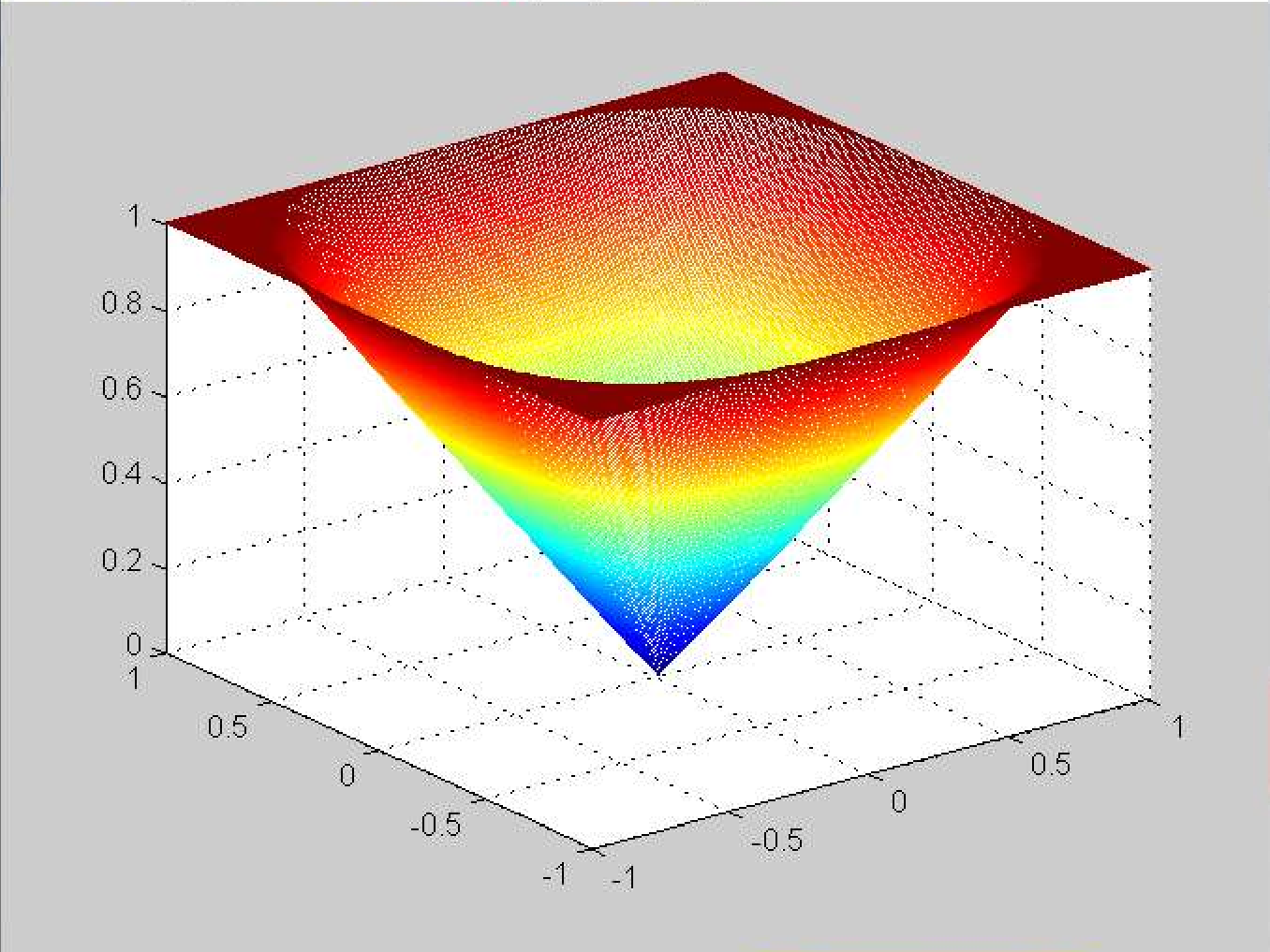}
  \label{fig_422}}}
  \vspace{0.2cm}

 \caption{Monogamy relation of different quantum measures with different dimension dependency properties in 3 qubit pure state system $\psi_{ABC}$. The surfaces are the boundary surfaces between monogamous and non-monogamous states w.r.t. correspondent quantum measures. The x and y-axes are the concurrences of $\rho_{AB}$ and $\rho_{AC}$, and z-axis is the concurrence of the bipartite system $A:BC$. Here we extend the range of x and y axes from [0,1] to [-1,1] for a better visual intuition. The monogamous/non-monogamous states are the configurations above/below the boundary surfaces. It can be observed that a bigger $s_{Q}(2)$ and a smaller $s_{Q}(3)$ result in a bigger volume of monogamous states}\label{fig_all}
\end{figure}

\begin{figure}
  \centering
  \includegraphics[width=12cm]{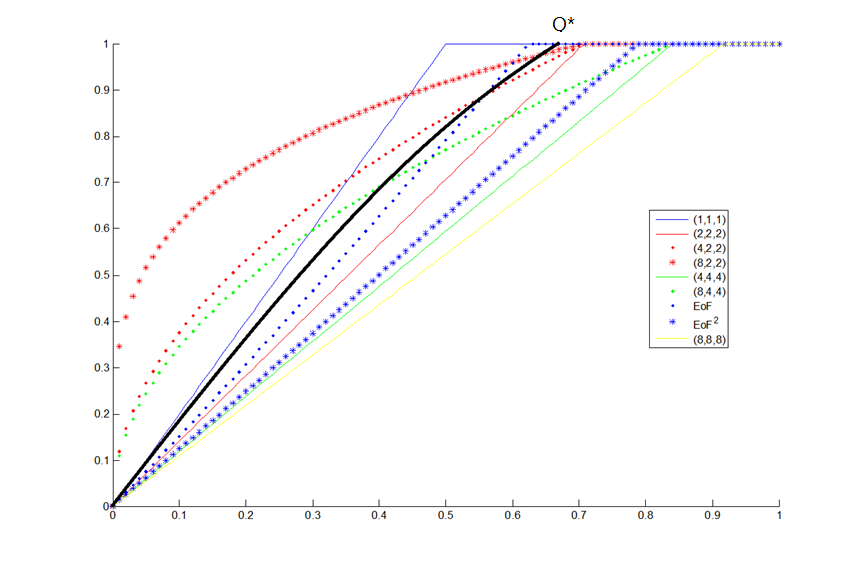}
  \caption{Higher powers of correlation measures and monogamy in 3 qubit pure state systems. Under our simplified assumption of the dimension dependency of quantum measures, the monogamy relation is given by $\mathcal{C}(1:2...n)^{s_{Q}(n)}\geq \sum_{i=2}^{n} \mathcal{C}((1:i))^{s_{Q}(2)}$ and in this figure it's abbreviated as $(s_{Q}(3),s_{Q}(2),s_{Q}(2))$. The curves show typical 2D section of the boundary surfaces of different quantum measures as given in Fig. []. The genuine measure for entanglement monogamy relation (possibly the squashed entanglement) is quantitatively given by the black curve such that $\mathcal{C}$ and $E_{F}$ do not satisfy the monogamy relation but $\tau=\mathcal{C}^{2}$ and $E_{F}^{2}$ do.}\label{fig_power}
\end{figure}

To give a first impression on how the dimension dependency will affect the monogamy relation, in Fig. \ref{fig_all} we show  the boundaries between monogamous and non-monogamous states with different measures for 3-qubit pure states. Fig. \ref{fig_power} shows a typical 2-dimensional intersections of the state space shown Fig. \ref{fig_dependency} of different measures. We can easily observe that an increased $s_{Q}(2)$ and decreased $s_{Q}(n)$, the relative volume of monogamous states will increase.

To verify that any quantum measure $Q$ can be made monogamous by using its higher power version $Q^{m},m>1$, from the geometrical point of view, we need to show that any valid entangled states satisfying the monogamy relation of $Q^{*}$ should fall in the monogamous region of $Q^{m}$ with a big enough $m$.

Obviously the genuine measure $Q^{*}$ for the monogamy property should have a boundary as given in Fig. \ref{fig_power} , which means that some of the monogamous states defined by $Q^{*}$ can not satisfy the monogamy relation for $\mathcal{C}$ and $E_{F}$ but they are all monogamous for $\tau$ and $E_{F}^{2}$. Also it can be easily observed that all the monogamous states of a given measure $Q$ can satisfy the monogamy relation of $Q^{m},m>1$. With the increase of $m$, the monogamous volume of $Q^{m}$ will be enlarged so that finally for a high enough $m$, the monogamous volume of $Q^{*}$ becomes a subset of the monogamous volume of $Q^{m}$ and therefore all entangled states are monogamous for the measure $Q^{m}$. For quantum measures that do not show such a simple dimension dependency with $\mathcal{C}$ such as the negativity as indicated in Fig. \ref{fig_dependency}, we have $s_{Neg}(3)=1$ and $s_{Neg}(2)>s_{C}(2)$. Since $\mathcal{C}^{2}$ satisfies the monogamy relation for 3-qubit pure states, therefore squared negativity does too.

 This idea can also be extended to multiple qubit systems and higher dimensional cases and thus gives a geometrical explanation of that all entanglement measures can be made monogamous by considering their higher power versions. Also we see that such monogamous relation does not reveal the geometrical nature of entanglement monogamy since the genuine measure that encodes the underlying geometrical nature of monogamy relation should be $Q^{*}$.

\subsection{Q2: Do large number of parties enforce monogamy in all quantum correlations?}
In \cite{Kumar_monogamy_qubit} numerical simulation on multiple qubit pure states shows that for any quantum measure, the percentage of states that fulfill the monogamy inequality increases with the number of parties. \cite{Kumar_monogamy_qubit} then concludes that the volume of the monogamous pure quantum states increases with an increasing number of parties.

From our geometrical point of view, the volume of the monogamous pure quantum states does not increase with the increasing number of parties since we assume that all the entangled states satisfy the monogamy of entanglement defined by the genuine measure $Q^{*}$. The correct explanation of the numerical results of \cite{Kumar_monogamy_qubit} is that the percentage of states that fulfill the monogamy inequality using any geometric measure $Q$ increases with the number of parties. We will show that it's the dimension dependencies of the genuine measure $Q^{*}$ and some measure $Q$ that leads to this observation.

Similar to our previous discussion, still we need to choose a proper reference for the dimension dependency. But since we are now considering the monogamy relation on systems with different number of parties, the dimension dependency of $\mathcal{C}$ is generally unknown. So $\mathcal{C}$ is not a proper reference. We will go back to \ref{eq_monogamy_simplified} and define the dimension dependency using a reference $\varepsilon$, which is unknown but will not affect our discussion as shown below.

Depending on the dimension dependency of a general quantum measure $Q$ and the genuine measure $Q^{*}$, for a multiple qubit pure state system we can classify our problem in the following cases:

\begin{description}
  \item[Case a] $Q$ is dimension independent, $s_{Q}(n)=s_{Q}(2)=const$ and $s_{Q}(2)< s_{Q^{*}}(2)$. 
  \item[Case b] $Q$ is dimension independent, $s_{Q}(n)=s_{Q}(2)=const$ and $s_{Q}(2)\geq s_{Q^{*}}(2)$. 
  \item[Case c] $Q$ is dimension dependent and $s_{Q}(2)< s_{Q^{*}}(2), ds_{Q}(n)/dn< ds_{Q^{*}}(n)/dn$. $E_{F}$, $\mathcal{C}$ and negativity seem to fall in this category even without a concrete proof.
  \item[Case d] $Q$ is dimension dependent and $s_{Q}(2)\geq s_{Q^{*}}(2), ds_{Q}(n)/dn< ds_{Q^{*}}(n)/dn$. Accordingly $E_{F}$, $\mathcal{C}^{2}$ and squared negativity seem to be an example of this category.
  \item[Case e] $Q$ is dimension dependent and $s_{Q}(2)< s_{Q^{*}}(2), ds_{Q}(n)/dn \geq ds_{Q^{*}}(n)/dn$. We do not know which existing measure belongs to this case yet.
  \item[Case f] $Q$ is dimension dependent and $s_{Q}(2)\geq s_{Q^{*}}(2), ds_{Q}(n)/dn \geq ds_{Q^{*}}(n)/dn$. Also we do not have an example of this type of measure yet.
\end{description}

\begin{figure}[tbp]
  \centering
  \mbox{
  \subfigure[$s_{Q}(n)=s_{Q}(2)=const,s_{Q}(2)< s_{Q^{*}}(2)$]
  {\includegraphics[width=6cm]{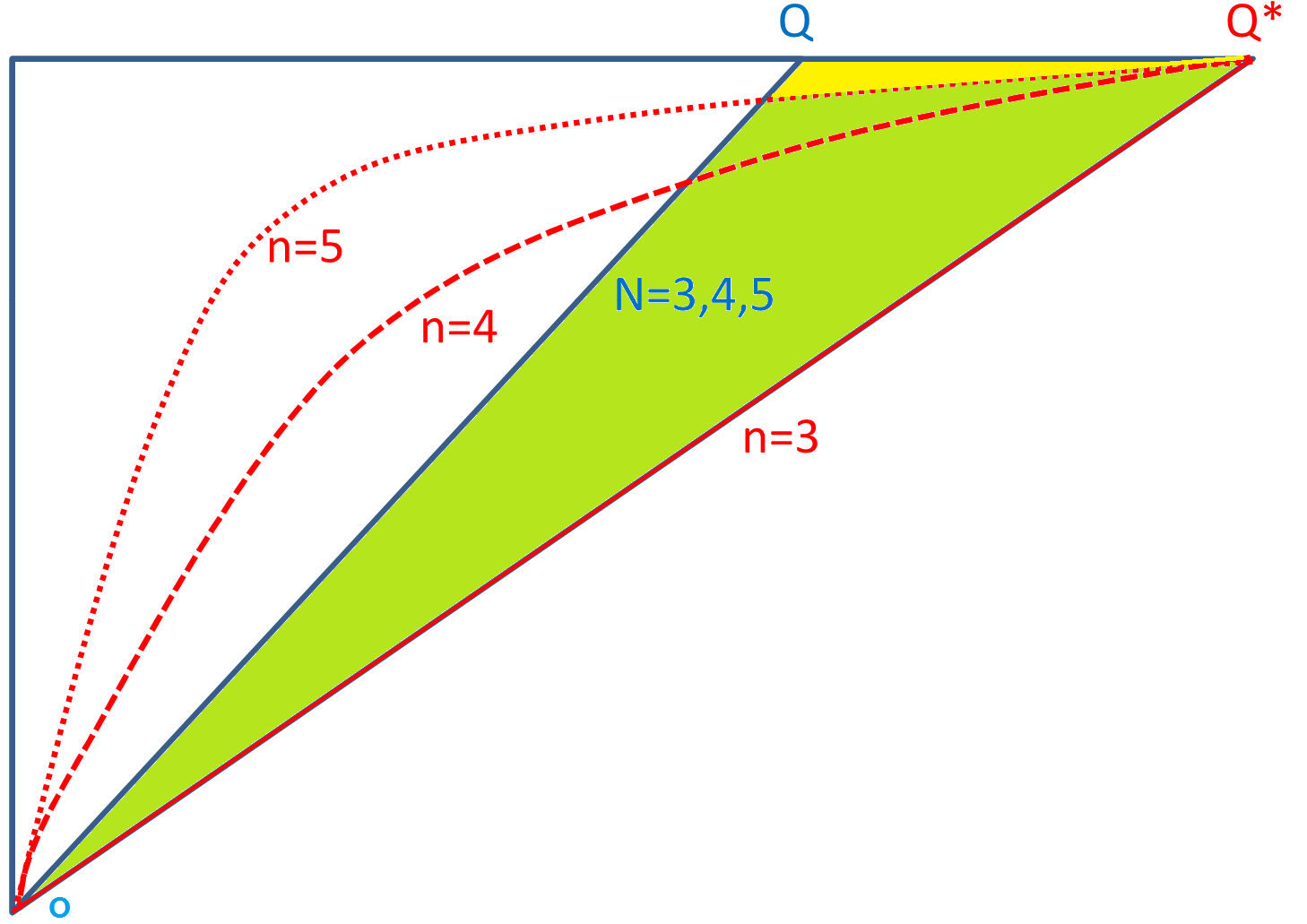}
  \label{fig_case1}}

  \subfigure[$s_{Q}(n)=s_{Q}(2)=const,s_{Q}(2)> s_{Q^{*}}(2)$]{
    \includegraphics[width=6cm]{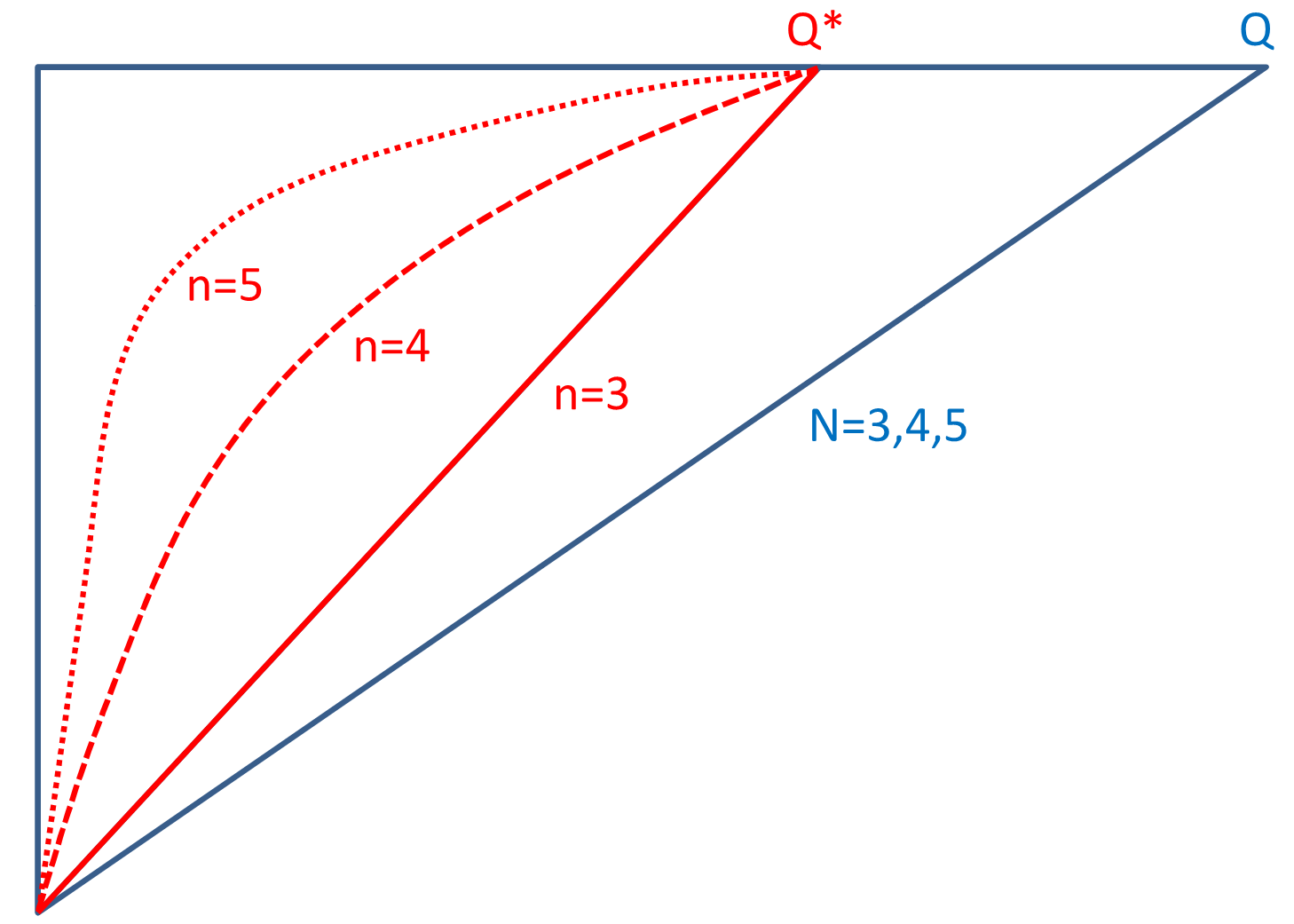}
  \label{fig_case2}}}
  \vspace{0.2cm}

\mbox{
   \subfigure[$s_{Q}(2)<s_{Q^{*}}(2), ds_{Q}(n)/dn< ds_{Q^{*}}(n)/dn$]{
  \includegraphics[width=6cm]{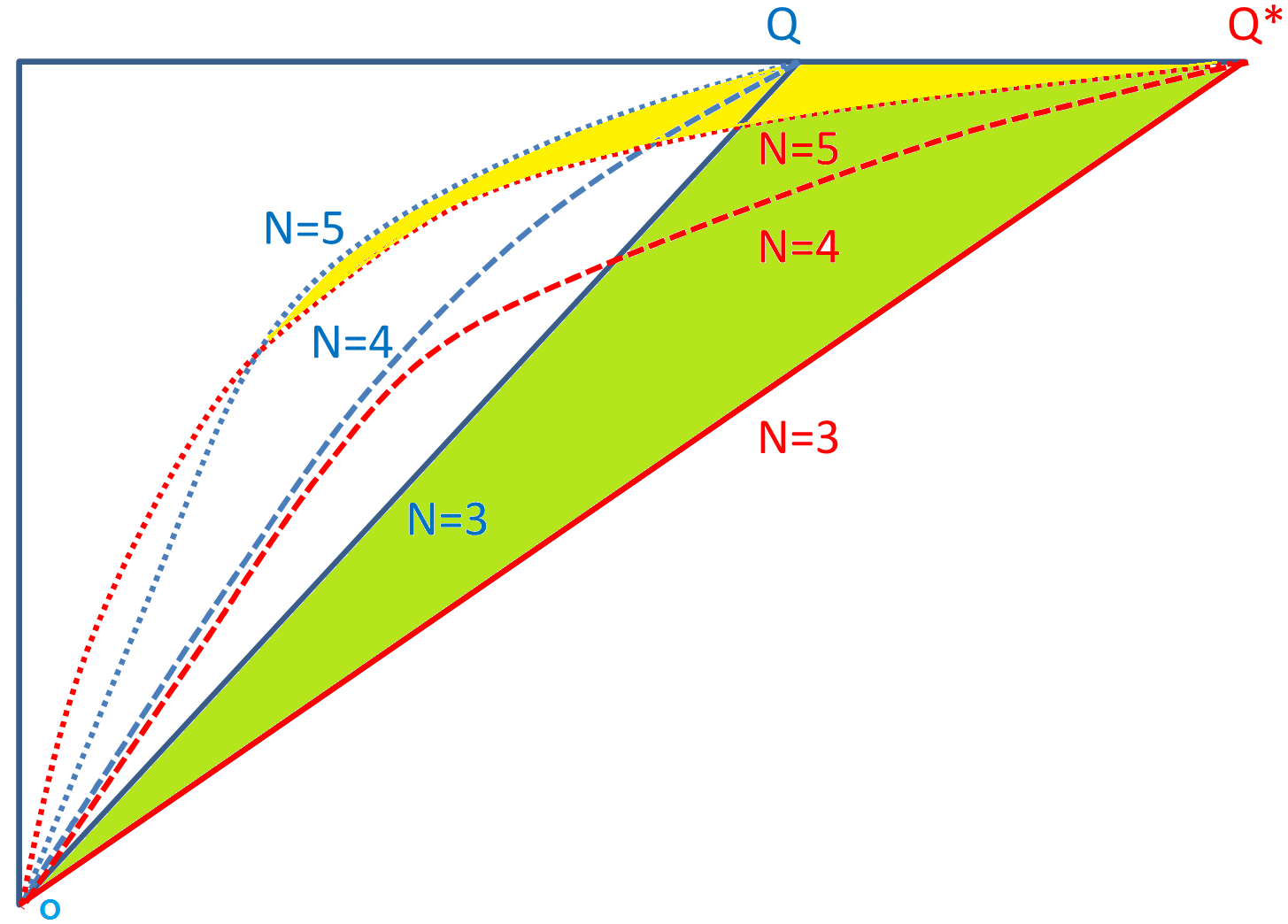}
  \label{fig_case3}}

   \subfigure[$s_{Q}(2)\geq s_{Q^{*}}(2), ds_{Q}(n)/dn< ds_{Q^{*}}(n)/dn$]{
    \includegraphics[width=6cm]{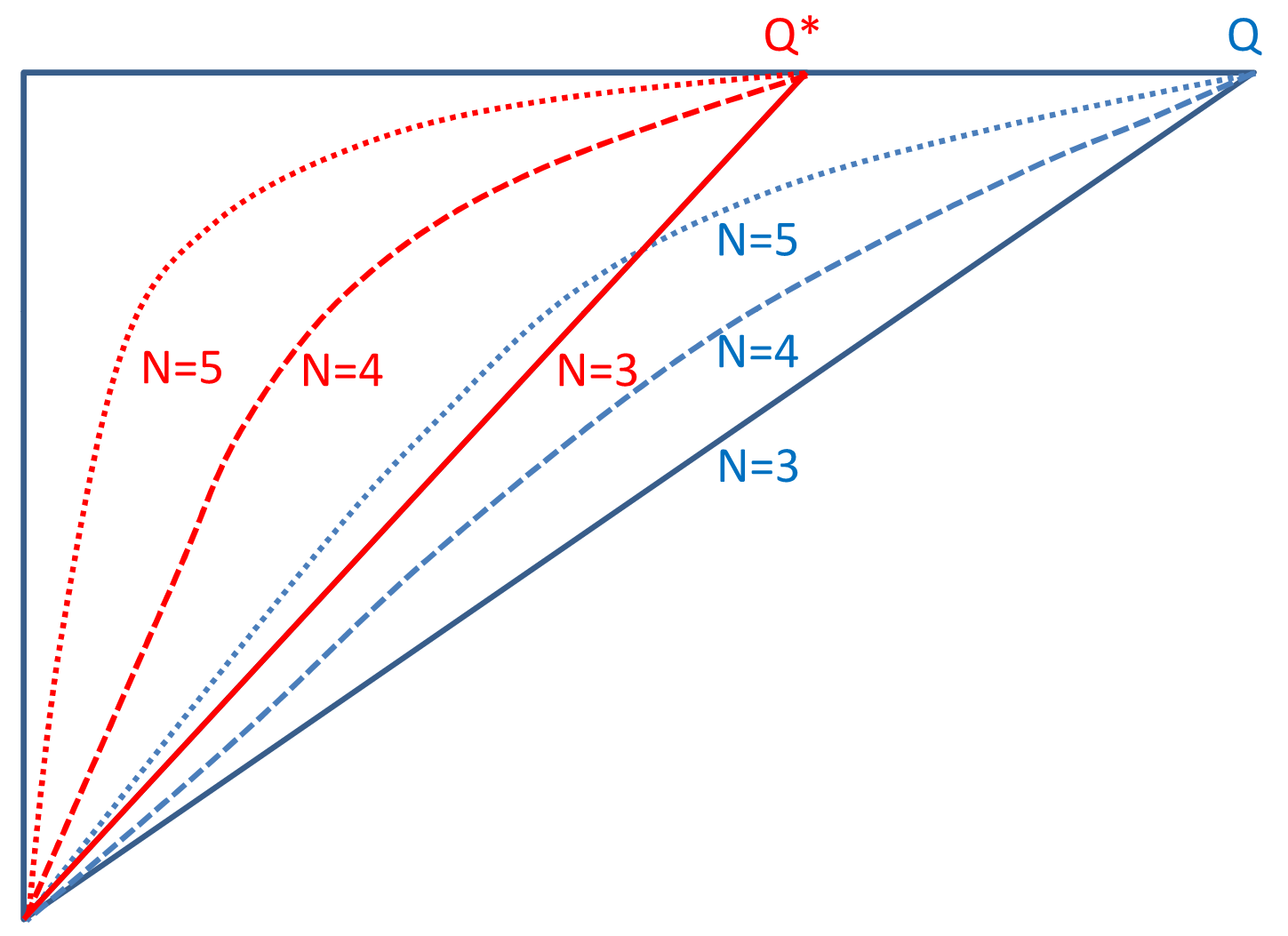}
  \label{fig_case4}}}
  \vspace{0.2cm}

  \mbox{
   \subfigure[$s_{Q}(2)< s_{Q^{*}}(2), ds_{Q}(n)/dn \geq ds_{Q^{*}}(n)/dn$]{
    \includegraphics[width=6cm]{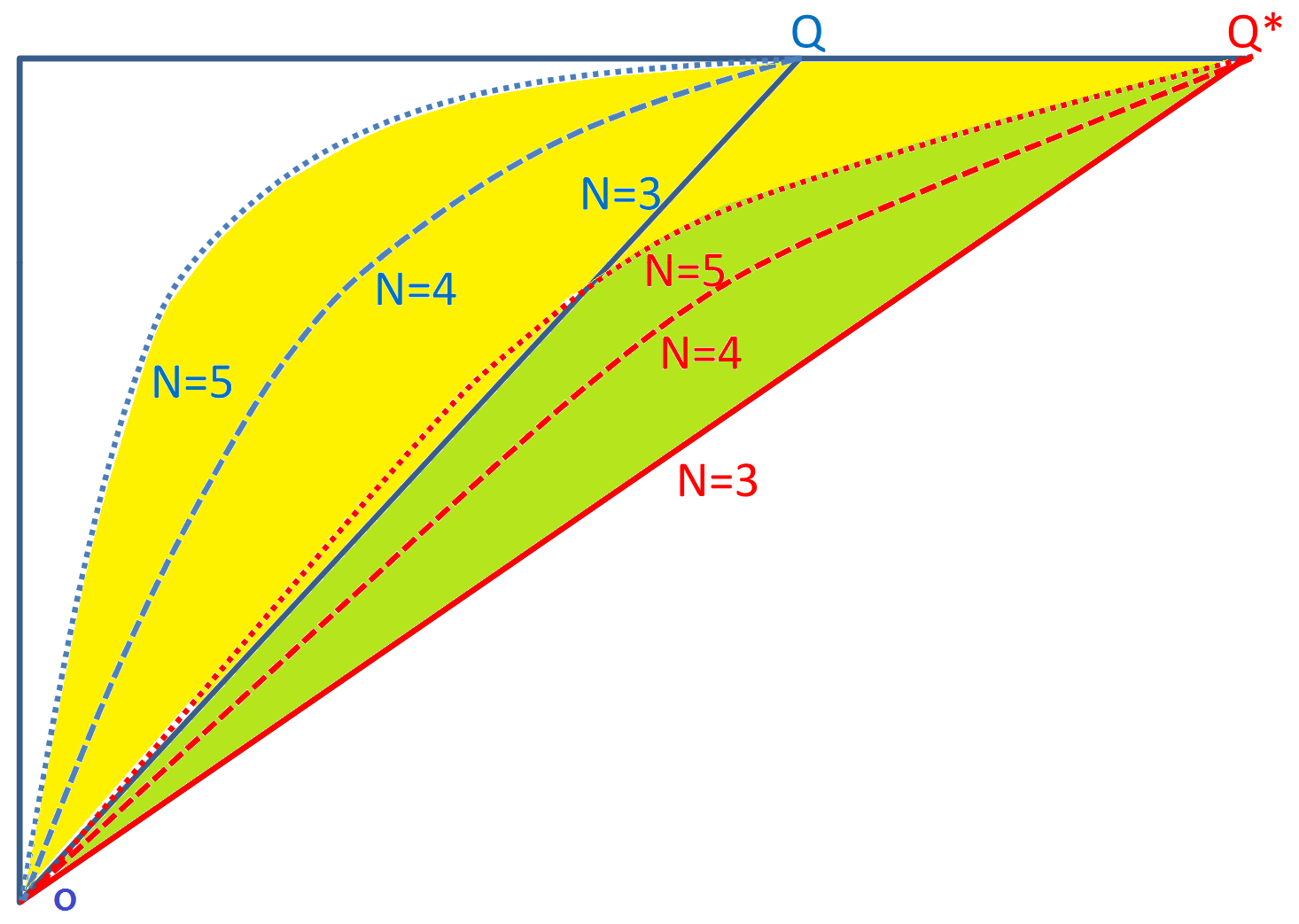}
  \label{fig_case5}}

   \subfigure[$s_{Q}(2)\geq s_{Q^{*}}(2), ds_{Q}(n)/dn \geq ds_{Q^{*}}(n)/dn$]{
    \includegraphics[width=6cm]{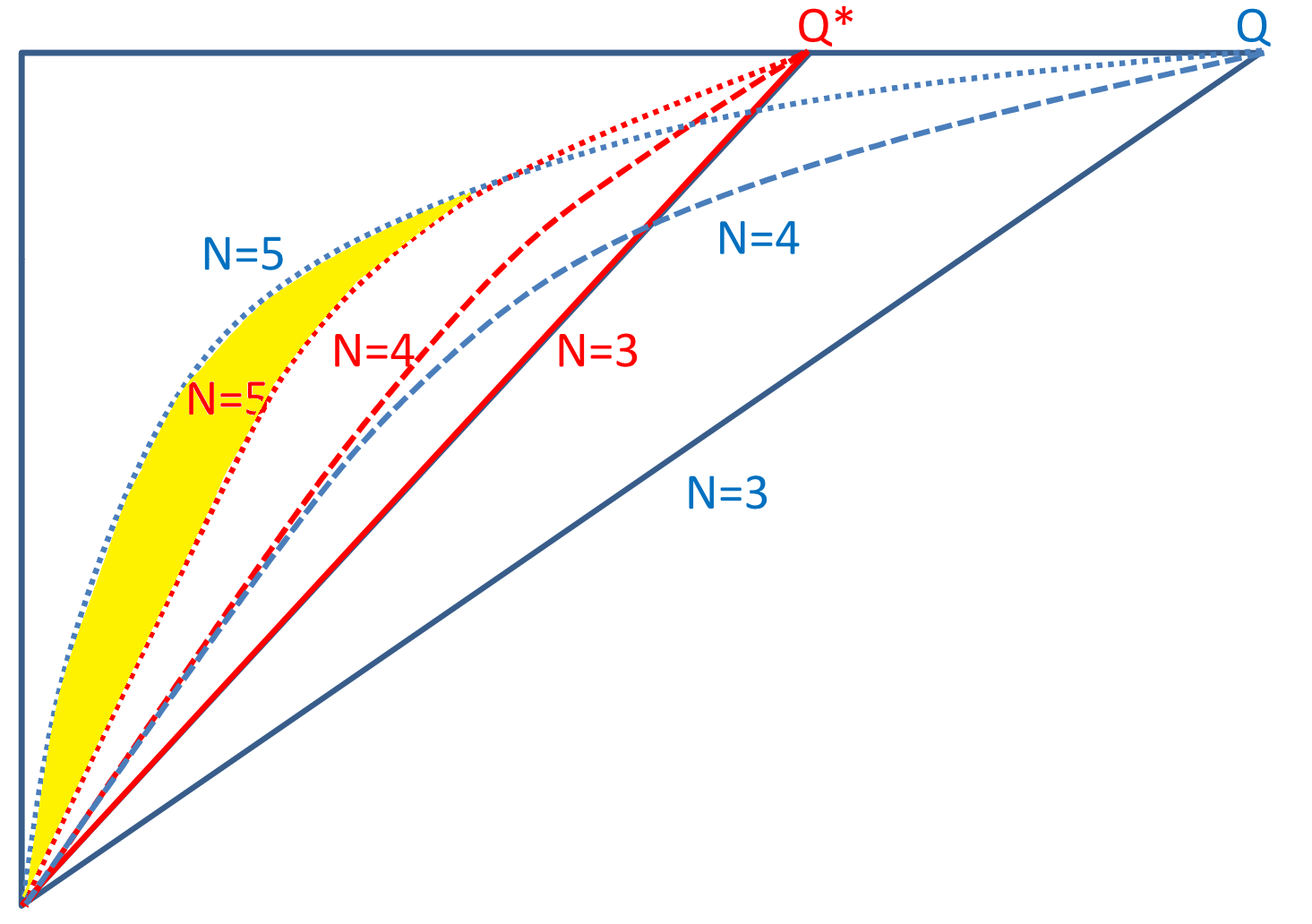}
  \label{fig_case6}}
  }
 \caption{Relation between monogamy of entanglement and dimension dependency of different entanglement measures with an increase of the number of qubits N. States that satisfy the monogamy relations of $Q$ and $Q^{*}$ fall in the left-upper part of the correspondent boundaries and all states fulfill the monogamy relation defined by $Q^{*}$. For cases (a)(c), the volume of states that are monogamous with $Q^{*}$ but non-monogamous with $Q$ decreases with a larger n from green region to yellow region. Also the size of yellow region decay to 0 with n increases; for cases(b)(d), all valid states fulfilling monogamy of $Q^{*}$ are also monogamous with $Q$; in cases (e), volume of states that are monogamous with $Q^{*}$ but non-monogamous with $Q$ does not decay to zero with n (from the triangle $OQQ*$ for $n=3$ to the yellow region when $n=5$); in case (f), the volume of non-monogamous states with $Q$ may even increase with $n$ as given by the yellow region for $n=5$.} \label{fig_all_cases}
\end{figure}

We then check how the monogamy relation evolves in different cases with an increase number of qubits. Remember that the genuine measure $Q^{*}$ determines the volume of valid entangled states since all entangled states are monogamous with $Q^{*}$ by our assumption.
Fig. \ref{fig_all_cases} shows the typical 2D sections of the boundary surfaces of the monogamy relation in all cases. The volume of non-monogamous states with $Q$ is then roughly proportional to the volume of the states that fall in the right side of the boundary of $Q$ and the left side of the boundary of $Q^{*}$ in each figure if the non-uniform distribution of the states are considered. It can be observed that:

(1) For cases (a)(c) where concurrence $\mathcal{C}$ and $E_{F}$ as typical examples, with the increase of the number of qubits, more states fulfilling the monogamous relation of $Q^{*}$ fall in the monogamous region of the measure $Q$ (from the triangle $OQQ^{*}$ when $n=3$ to the yellow region when $n=5$). This is exactly what's observed by \cite{Kumar_monogamy_qubit}. Although it can be observed that with a larger qubit number the volume of non-monogamous states w.r.t. $Q$ decrease to zero approximately, it's a reasonable guess that the set of non-monogamous states w.r.t. $Q$ should not be a zero-measure set even for pure states. \cite{Kumar_monogamy_qubit} mentioned a similar scenario for mixed states that symmetric mixed states form a non-zero, perhaps fast-decaying, volume of
monogamous multiparty quantum states of all quantum states, for large systems. From Fig. \ref{fig_case1}\ref{fig_case3} we think it's highly possible that for pure states the non-monogamous multiparty states w.r.t. $Q$ has a fast-decaying but non-zero volume. We think that the reason that that the possible non-zero volume of non-monogamous states is not observed in \cite{Kumar_monogamy_qubit} is that the volume decays fast with $n$ and the numerical simulation has a limited accuracy.

(2) For cases (b)(d), all the states satisfy the monogamous relation of $Q$. They are correspondent to the cases of tangle $\tau$ for all qubit systems and $E_{F}^{2}$ or squared negativity for 3 qubits as described in \cite{Kumar_monogamy_qubit}.

(3) Case (e) shows an interesting property that there are always a set of states with non-zero measure that does not fulfill the monogamous relation w.r.t. $Q$. Also the relative volume of this set can even increase with an increase of the number of qubits. So far such phenomenon has not been observed, but we can not rule out such a possibility.

(4) Case (f) is even more wierd. According to the figure, we can find that when $n=3$ all states are monogamous w.r.t. $Q$, but non-monogamous states may appear with an increased $n$ as shown by the yellow region in \ref{fig_case6} when $n=5$. To the best of our knowledge, if such kind of entanglement measure exists is an open question. But theoretically we can not exclude its existence.

So the increase of relative volume of monogamous states w.r.t. some entanglement measures is due to the dimension dependency of the quantum measures. But it may not hold as a general property of all entangled states and entanglement measures. It might be an interesting problem to find out if there exist quantum measures with the dimension dependency properties of cases (e)(f).

\section{Discussions}
Our work is heavily based on several basic assumptions, including
\begin{itemize}
  \item We regard the  nature of entanglement as a geometrical structure and assume all the properties of entanglement, including entanglement measures and the intrinsic monogamy of entanglement, are emergent from this structure.
  \item Generally the geometrical structure is dimension dependent and accordingly entanglement measures,the genuine measure for the intrinsic monogamy of entanglement $Q^{*}$ , will also be dimension dependent.
\end{itemize}

In this part we will try to clarify our understanding of the above problems, i.e., what might be the geometrical structure of entanglement and why $Q^{*}$ is dimension dependent.

\subsection{Geometry of entanglement}
Here we will address two different approaches to understand the geometry of entanglement. The first idea is to regard the entanglement as the twisting of a nontrivial fibre bundle structure on the state space. Roughly speaking, for an entangled composite system with subsystems A and B, a nontrivial fibre bundle can be constructed with the state space of one of the subsystems as the base space and the other subspace as the fibre. A proper connection is defined on the fibre bundle and the fibres of different points on the base space are related by the parallel translation determined by the connection. The entanglement of the two subsystems is then modeled by this parallel translation operation,i.e., a change of the configuration of the base space, subsystem A, will lead to a change of configuration of the fibre space, subsystem B. This idea has been successfully demonstrated on 2 qubit pure state systems quantitatively. Though effort to extend this idea to more complex systems such as 3 and 4 qubit systems are not so successful, we think it's a mathematically elegant candidate for the geometry of entanglement. To extend this approach to understand the entanglement in mixed states or to multipartite systems will be very interesting.

A more fruitful approach is the famous ER=EPR conjecture of \cite{Maldacena_ER_EPR}\cite{Susskind_ER_EPR}\cite{Susskind_blackhole}\cite{Susskind_ER_bridge}\cite{Susskind_ER_bridge_nowhere}, which tries to understand entanglement by identifying the entanglement between subsystems with the ER bridge connecting them. Though it still invites lots of debate, it has been explored to understand the AMPS paradox of black hole information and it's shown that the geometry of ER bridge is closely related with quantum information concepts\cite{Susskind_ER_bridge}\cite{Susskind_ER_bridge_nowhere}.

But what does it mean by the elegant formula $ER=EPR$? There might exist several possible interpretations, including
\begin{itemize}
  \item (a) One of them is more fundamental, i.e., either $EPR\Rightarrow ER$ or $ER\Rightarrow EPR$. Here $A\Rightarrow B$ mean B is generated by A so that A is more fundamental.
  \item (b) ER and EPR are dual to each other so that there exists a one-to-one correspondence between their state space.
  \item (c) Both ER and EPR are emergent from the same root, for example an underlying geometrical structure. But there is no direct correspondence between ER and EPR.
\end{itemize}

 For the first assumption, currently there are some works indicating that ER can be emergent from EPR, i.e., ER bridges or microscopic wormhole structures exist in certain entangled systems\cite{Hensen_EPR_wormhole}\cite{Lobo_EPR_wormhole}. But also they show EPR does not necessarily imply ER\cite{Hensen_EPR_wormhole}. This is consist with the declaration of \cite{Maldacena_ER_EPR} that the appearance of ER means that it must be generated by EPR but the converse statement is less certain. This is a sign that EPR maybe more fundamental than ER.

 For case (b), if ER and EPR are dual to each other, then they must have the same state space volume. So for different patterns of entanglement there should be their correspondent different ER bridges. \cite{Maldacena_ER_EPR} shows the correspondence between EPR and ER patterns  in entangled AdS black holes. \cite{Susskind_ER_EPR} discussed the consistency between ER=EPR and the quantum measurement by the property of multiple black holes entangled in the GHZ pattern. These works reveal interesting relation between entanglement patterns and ER bridges. But an quantitative description of how the ER bridges are generated from different entanglement patterns is still missing. For example, \cite{Susskind_ER_EPR} discussed the tripartite ER bridge in GHZ entanglement and indicated the mysterious GHZ-brane\cite{Susskind_ER_bridge} in the core of the ER bridge. But the description of \cite{Susskind_ER_EPR} is far from complete. How does the GHZ-brane forbids any two observers to communicate by just jumping into two black holes? What's the mechanism of the cutting-off of the ER bridge by the trace-out operation on one of the 3 black holes? What's the geometry of ER bridges corresponding to mixed states? How does the ER bridge geometry evolve local unitary operations? To fully explore the connection between EPR and ER, at least a detailed analytical explanation for the geometry of the GHZ-brane is needed and the ER bridge based picture does not seem to be able to achieve this goal.

We speculate maybe the solution is in case (c) so that there exists an underlying geometrical structure for both EPR and ER. Currently the only candidate for such a geometrical structure is the fibre bundle structure on the state space of quantum systems. To accomplish a complete geometrical description of entanglement in this approach, the forth-coming tasks are
\begin{itemize}
  \item Extending the work of the Hopf fibration on 2 qubit pure state systems to construct the fibre bundle structures and define correspondent proper connections for different systems, so that different entanglement patterns are described either as different fibre bundle structures or as inequivalent connections.
  \item Extending this approach to describe the entanglement of mixed states. Till now there is no concrete work on describing the geometrical structure of mixed states, even for 2 qubit mixed states. According to \cite{Fortin_partial_trace}, the density matrix for a mixed state does not assign any \emph{state} to the system in the same way as it does on a pure state system. So if the entanglement for mixed states has a real geometrical structure is still open. According to the ER=EPR conjecture, the geometry of ER bridges are correspondent to the pattern of entanglement. Obviously the entanglement pattern of a mixed state is encoded in its specific purification, so the entanglement of mixed states may not hold a concrete geometrical structure. On the counterpart, we should also be curious about how to describe the ER bridge geometry generated by mixed states and its relation with the current description of the geometry of mixed states\cite{Montgomery_geo_mixed_states}\cite{Andersson_geo_mixed_states}. For now, we speculate that both the ER bridge geometry of mixed states and the bridge-to-nowhere structure of a single black hole should possess a dynamic varying cross-sectional geometry.

  \item Understanding how this geometrical structure evolves with different operations including merging subsystems, unitary transformation and measurement. From the fibre bundle structure point of view, we can interpret the effects of these three operations as follows: The merging of subsystems provides a state space with a higher dimension to support more complex fibre bundle structures; unitary operations can build different fibre bundle structures on the state space and usually this operation need a scrambling time to achieve the goal as explained in \cite{Susskind_ER_EPR}; and measurement can separate systems into subsystems and also be capable to help to build geometrical structures as described in \cite{Maldacena_ER_EPR}.
\end{itemize}

So the above mentioned fibre bundle structure can be regarded as a possible microscopic description of entanglement and the ER bridge picture of \cite{Maldacena_ER_EPR} is a more intuitive microscopic picture. Combining these two pictures and taking the GHZ-state tripartite black hole system as an example, to build the underlying geometrical structure for it, we should answer the following questions
\begin{itemize}
  \item How to explain the GHZ-brane property by constructing fibre bundles on the state space of the system?
  \item How the fibre bundle structure can be built by the measurement operation from a Bell state 2 qubit system as given in \cite{Susskind_ER_EPR}\cite{Susskind_ER_bridge}?
  \item How the tracing out of one of the 3 parties will destroy the fibre bundle structure and result in an unentangled system?
\end{itemize}

\begin{figure}
  \centering
  \includegraphics[width=10cm]{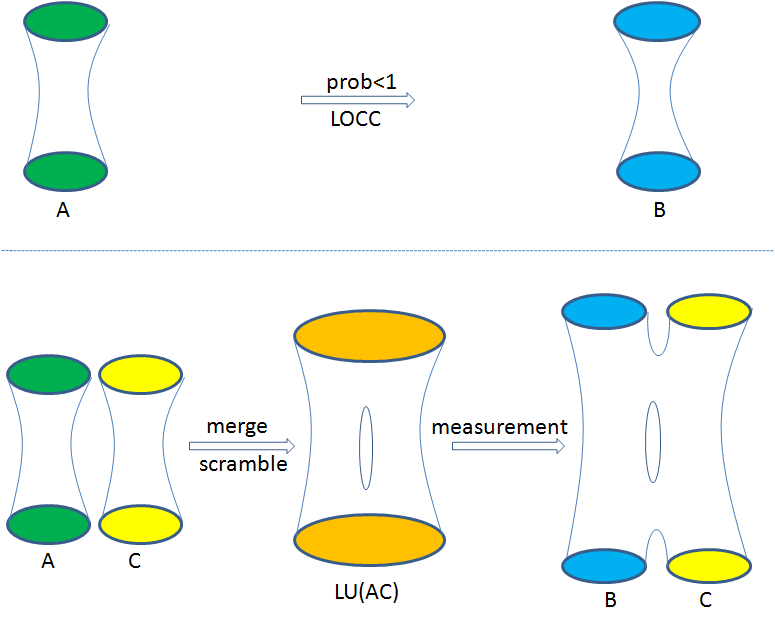}
  \caption{Geometric picture of the entanglement catalyst operation. Originally entanglement state A can not be transformed to state B with certainty; Introducing the catalyst system C followed by operations including emerging subsystem A and C, local unitary operation on composite system AC, finally we get system B and C with certainty. The key component here is that emerging subsystem A and C and local unitary operations on AC create new geometrical structures which were not possible with only system A, so that the transformation from AC to BC can be achieved with certainty.}\label{fig_geo_catalyst}
\end{figure}

\begin{figure}
  \centering
  \includegraphics[width=12cm]{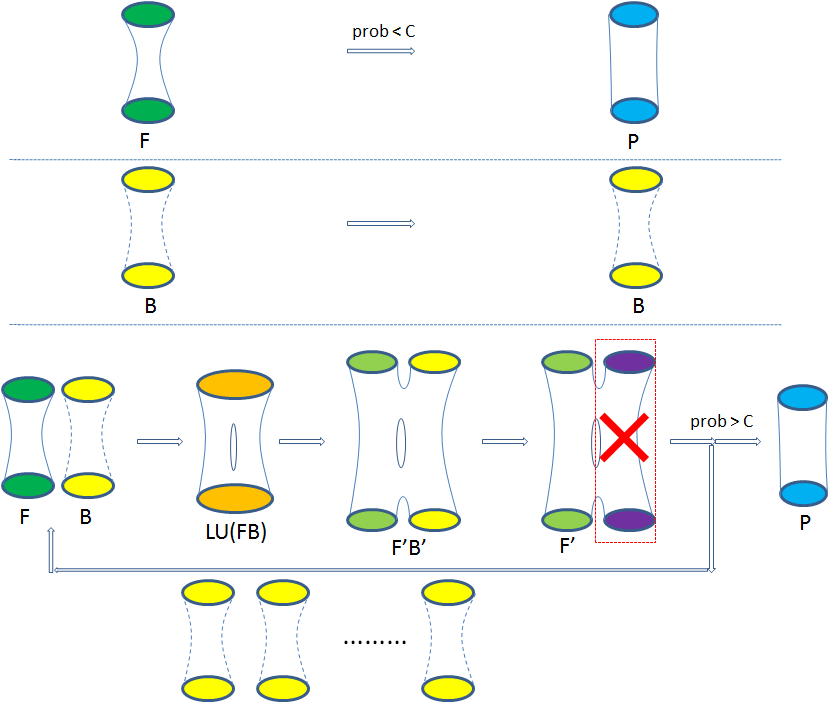}
  \caption{Geometric picture of the bound entanglement activation. Top: A free entangled state can only be distilled with a limited success probability C; Middle: A bound entanglement state can not be distilled; Bottom: Combining the free entangled and a set of bound entangled states can achieve an arbitrary good success probability of distillation \cite{Horodecki_bound_activation}. }\label{fig_bound_activation}
\end{figure}

In \cite{Susskind_ER_EPR} also the ER bridge based macroscopic picture was used to explain the teleportation operation, where the system merging, unitary transformation and measurement play their roles on the evolution of the geometrical structures of the system as we described before. The key observation here is that \emph{a higher dimension provides more possibilities}. In fact this is also the key component for other quantum information processing tasks including the entanglement distillation, entanglement catalyst and the activation of bound entanglement, where the merging of subsystems plays the role to turn mission impossible to mission possible due to the emergence of new geometrical structures by merging subsystems and scrambling operations. Accordingly to clarify how the geometrical structures evolve in these operation protocols should also appear on our mission list. Fig. \ref{fig_geo_catalyst} and Fig. \ref{fig_bound_activation} show examples of the geometric pictures of the entanglement catalyst  and bound entanglement activation operations respectively. Of course this is  far from the end of the story since we need to figure out the mathematical details of how the geometrical structures of the system, for example the fibre bundle structure, change during these operations.

\subsection{Geometry of entanglement monogamy}
There are adequate evidences that entanglement and entropy are closely related to geometry. Also generally entropy inequalities admit geometrical explanations\cite{Maldacena_ER_EPR}\cite{Nishiokaa_HEE}. It's very unlikely that the monogamy of entanglement is an exception.

Why general monogamy relation is unique for entanglement and geometrical? Following the proof of \cite{Streltsov_monogamy_general}, based on a few basic assumptions including (a) positivity; (b) invariance under local unitary transformations; and (c) nonincreasing when an ancilla is introduced, the result was drawn that general monogamy relation does not hold in general for a correlation measure that does not vanish on separable states. Obviously discord also fulfills the constraints. It's exactly the general monogamy condition of $Q$ on \emph{any} system that excludes the discord as a monogamous measure. Also we know that monogamy of entanglement holds for squashed entanglement. Therefore we can conclude that general monogamy relation is an intrinsic property of entanglement.

A further check of the general monogamy relation, $Q(A|BC)>Q(A|B)+Q(A|C)$, shows that it actually can be understood as follow: For any system $\rho_{AB}$ and its arbitrary extension $\rho_{ABC}$ with $Tr_{C}\rho_{ABC}=\rho_{AB}$, the monogamy relation holds. Physically this means that the general monogamy relation is essentially a property of all the open systems $\rho_{ABC}$ taking $\rho_{AB}$ as its subsystem. So the genuine entanglement measure $Q^{*}(A|B)$ should be defined on all the extensions $\rho_{ABE}$ of $\rho_{AB}$. This is exactly what the squashed entanglement does. From this interpretation of the general monogamy relation, we notice that the proof of [] is a special case of it, where it focuses on a separable state $\rho_{AB}=\sum_{i}p_{i}|\psi_{i}>_{A}<\psi_{i}|\otimes |\psi_{i}>_{B}<\psi_{i}|$ and a special extension of it as $\rho_{AB}=\sum_{i}p_{i}|\psi_{i}>_{A}<\psi_{i}|\otimes |\psi_{i}>_{B}<\psi_{i}|\otimes |\psi_{i}>_{C}<\psi_{i}|$ with orthogonal states $\{|\psi_{i}>_{B}\}$. The general monogamy relation on this special case leads to the conclusion that $Q(A|B)=0$ for any separable state $\rho_{AB}$. Why doesn't the entanglement of formation $E_{F}$ fulfill the general monogamy relation? Recalling for a system $\rho_{AB}=\sum_{i}p_{i}|\psi_{i}>_{B}<\psi_{i}|$ the definition of the squashed entanglement as $E_{sq}(\rho_{AB})=inf_{E}\frac{1}{2}I(A:B|E),\rho_{AB}=Tr_{E}\rho_{ABE}$ and $E_{F}(\rho_{AB})=inf_{E}\frac{1}{2}I(A:B|E),\rho_{ABE}=\sum_{i}p_{i}|\psi_{i}>_{B}<\psi_{i}|\otimes |\psi_{i}>_{E}<\psi_i|$ with orthogonal states ${|\psi_{i}>_{E}}$. A heuristic interpretation of the non-monogamy of $E_{F}$ is that $E_{F}$ is a special version of the squashed entanglement since for $E_{F}$ the infinum is taken on a special subset of all possible extensions of $\rho_{AB}$. The lack of exploring the whole extension space makes $E_{F}$ not a general monogamous measure. Of course we can not claim that the squashed entanglement is the unique genuine measure for the general monogamy of entanglement since there might be other general monogamous entanglement measures.

\begin{figure}
  \centering
  \includegraphics[width=12cm]{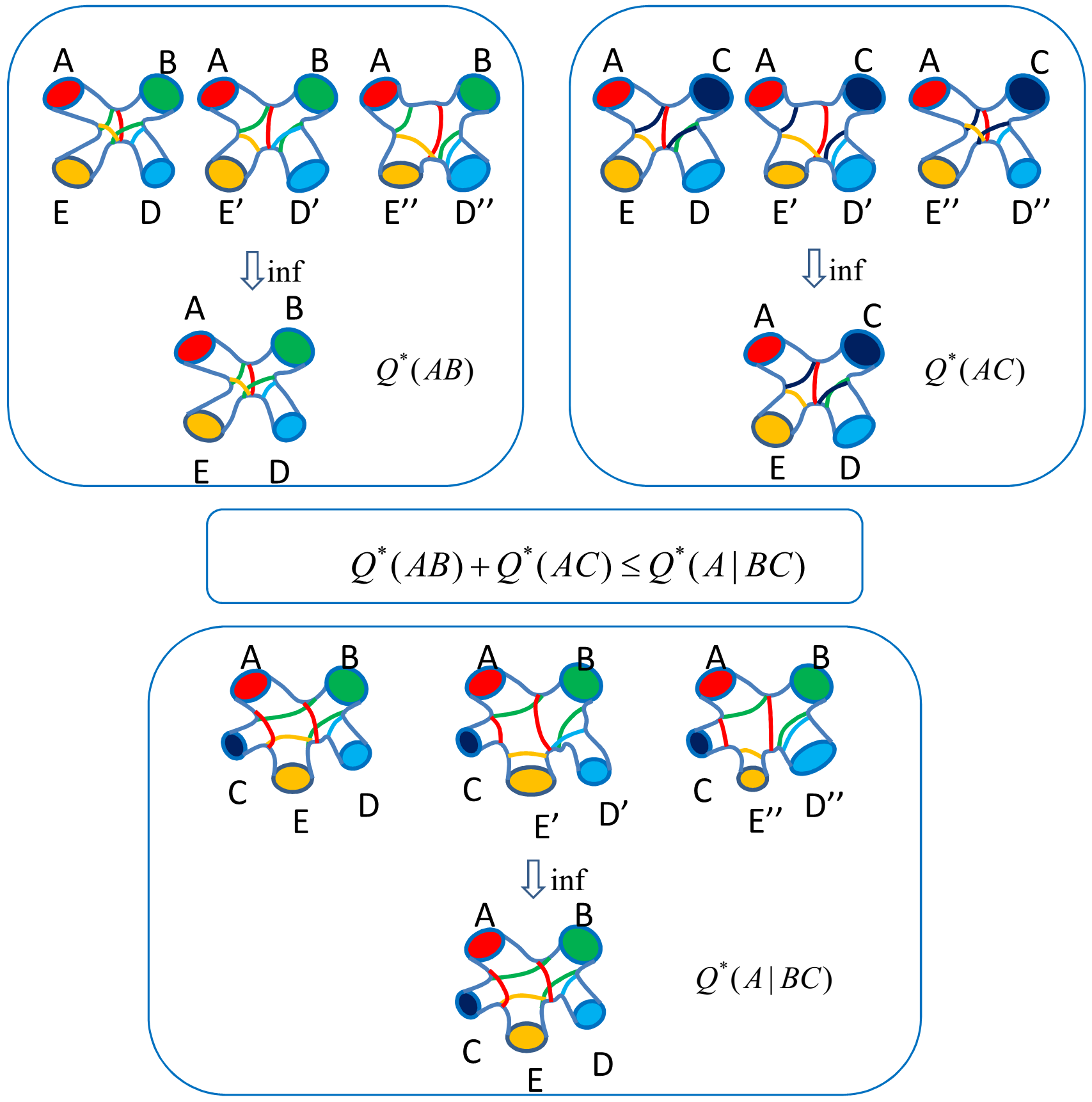}
  \caption{Geometry of monogamy of entanglement w.r.t. squashed entanglement from the ER=EPR conjecture. Left top: The squashed entanglement of a mixed state $\rho_{AB}$ with $E,E',E''$ as arbitrary extensions of $\rho_{AB}$ and $D,D',D''$ as the environment, the colored curves on the ER bridge represent the minimal surfaces as explained in \cite{Gharibyan_ER_EPR_inequation}; Right top: The squashed entanglement of $\rho_{AC}$; Bottom: The squashed entanglement of $A|BC$.}\label{fig_geo_monogamy}
\end{figure}

The geometry of monogamy relation can be understood by the language of $ER=EPR$ conjecture. In $ER=EPR$, the entanglement entropy is correspondent to the cross-sectional area of the ER bridge. The squashed entanglement which is built on entropy can then be represented by the geometries of the ER bridges determined by the system $\rho_{ABC}$ as shown in Fig. \ref{fig_geo_monogamy}.

The above discussion shows that the monogamy of entanglement does have a geometrical interpretation and the genuine measure for entanglement monogamy is dimension dependent. It seems the later declaration is wrong since the ER bridge is in a space with limited dimension as shown in Fig. \ref{fig_geo_monogamy}. But if the conjecture $ER=EPR$ holds, then the pattern of an ER bridge is determined by the pattern of entanglement that creates it but not only on the amount of entanglement! So Fig. \ref{fig_geo_monogamy} is just an intuitive way to show that monogamy of entanglement has a geometrical picture, but the real situation should be more complex. Besides the obvious dimension dependency of the definition of the squashed entanglement, an example to show this is the tripartite ER bridge with a GHZ-brane in the center. The tracing out of any one of the tripartite, for example A, will destroy the GHZ-brane structure. So to compute the cross-sectional area with or without the mysterious GHZ-brane definitely have different complexities.

\section{Conclusions}
In conclusion, in this work we addressed the problem of how to understand the monogamy relation from the hypotheses that entanglement entanglement, entanglement measures and monogamy relation of entanglement are all regarded as emergent properties of a geometrical structure of the system state space. Based on a summary of the evidences that show the relationship between entanglement and geometry, we propose to understand the monogamy relations of different quantum measures by one key property of quantum measures, the dimension dependency. We then analysis the two observed facts about the monogamy of entanglement: why every correlation measure can be made monogamous by its higher power versions and why the volume of monogamous state seems increase with the increase of the number of subsystems? Detailed discussion about the possible answers of the geometry of entanglement and the geometric picture of general monogamy relation of correlation measures are also given. Although the geometry of entanglement and related concepts have attracted lots of research interest and led to promising results [][][][][], details on analytical quantitative results on how different properties of quantum entanglement related problems, such as inequivalent entanglement patterns, entanglement measures, entanglement distillation, bound entanglement and its activation, geometry of quantum computation, can be emergent from the underlying geometrical structure of the state space remains unknown. A complete picture of the geometry of entanglement will definitely enhance our understanding of physics problems far beyond quantum information processing.

\bibliographystyle{amsplain}
\bibliography{paperBib}
\end{document}